\documentclass[traditabstract]{aa}
\usepackage{psfrag,graphicx}
\usepackage{color}
\usepackage{txfonts}
\usepackage{natbib}
\usepackage{breqn}
\usepackage[T1]{fontenc}
\usepackage{ae,aecompl}

\title{The formation of supermassive black holes in rapidly rotating disks}
\titlerunning{Black hole  formation in high redshift protogalaxies}

\author{M.~A.~Latif \inst{1,2}
\and
D.~R.~G.~Schleicher \inst{3,4}
}
\institute{Sorbonne Universités, UPMC Univ Paris 06, UMR 7095, Institut d'Astrophysique de Paris, F-75014, Paris, France 
\and
CNRS, UMR 7095, Institut d'Astrophysique de Paris, F-75014, Paris, France
\and
Departamento de Astronomía, Facultad Ciencias Físicas y Matemáticas, Universidad de Concepción,  Av. Esteban Iturra s/n Barrio Universitario, Casilla 160-C, Chile 
\and
Scuola Normale Superiore, Piazza dei Cavalieri 7, 56126 Pisa, Italy
 }

\authorrunning{Latif \& Schleicher}
  
\date{}

\begin{document}
\bibliographystyle{aa}


%

%


 \begin{abstract}
{
Massive primordial halos exposed to moderate UV backgrounds are the potential birthplaces of supermassive black holes. In such a halo, an initially isothermal collapse will occur, leading to high accretion rates of $\sim0.1$~M$_\odot$~yr$^{-1}$. During the collapse, the gas in the interior will turn into a molecular state, and form an accretion disk due to the conservation of angular momentum. We consider here the structure of such an accretion disk and  the role of viscous heating in the presence of high accretion rates for a central star of $10$, $100$ and $10^4$~M$_\odot$. Our results show that the temperature in the disk increases considerably due to viscous heating, leading to a transition from the molecular to the atomic cooling phase. We found that the atomic cooling regime may extend out to several $100$~AU for a $10^4$~M$_\odot$ central star and provides substantial support to stabilize the disk. It therefore favors the formation of a massive central object. The comparison of clump migration and contraction time scales shows that stellar feedback from these clumps may occur during the later stages of the evolution. Overall, viscous heating provides an important pathway to obtain an atomic gas phase within the center of the halo, and helps in the formation of very massive objects. The latter may  collapse to form a massive black hole of  about $\geq 10^4$~M$_\odot$.

}
 \end{abstract}

\maketitle

\section{Introduction}

Self-gravitating accretion disks are common around astrophysical objects ranging from protoplanetary disks around stars to the disks in massive spiral galaxies as our  Milky-Way. They are formed as a consequence of  gravitational collapse due to the conservation of the angular momentum. The size of these disks varies from a few Astronomical Units (AU) to kiloparsec (kpc) scales. In the context of early structure formation, disks with sizes of a few hundred AU to pc scales are expected to form \citep{Abel2000,Bromm2002}. In fact, numerical simulations show the formation of such disks both in minihalos of $10^{5}-10^6$~M$_\odot$ \citep{Clark11,Greif12,LatifPopIII13} as well as in massive primordial halos of $\sim10^7$~M$_\odot$ \citep{Regan09,Latif2012,Latif2013c,Becerra2014} forming in the early universe at about $z=15$. The long term evolution and stability of such disks is however not well understood. Particularly, it has not been explored in the presence of a central massive object due to the computational costs required for a simulation pursuing a long time-evolution. 


The observations of high redshift quasars reveal that black holes of a billion solar masses were assembled within the first billion years after the Big Bang \citep{2003AJ....125.1649F,2006AJ....131.1203F,2011Natur.474..616M,2013ApJ...779...24V}.  The potential pathways for the formation of the earliest supermassive blacks holes include the seeds from stellar remnants, the direct collapse of a massive protogalactic gas cloud referred as the direct collapse and stellar dynamical processes in dense stellar clusters, see dedicated reviews for the details of these processes \citep{2010A&ARv..18..279V,Natarajan2011,2012arXiv1203.6075H}. The direct collapse scenario has emerged as the most feasible way to assemble massive black holes in a short time and is the focus of this work.  It mandates that gas must be of a primordial composition and the molecular hydrogen cooling remains suppressed.
The massive primordial halos of $\rm 10^7-10^8~M_{\odot}$ forming around $\rm z=15-20$ are the potential birthplaces of  the direct collapse black holes. It has been shown that supermassive stars of $\rm 10^4-10^6~M_{\odot} $ can form as a result of direct collapse in massive primordial halos if the cooling mainly proceeds via atomic lines \citep{Begelman2006,Volonteri2008,Latif2013d,Schleicher13,Hosokawa13}.  \cite{Ferrara14} have estimated the initial mass function of the resulting direct collapse black holes and found that previous UV feedback can play a central role in determining its shape. The molecular hydrogen cooling can be quenched in the ubiquity of  UV flux emitted from the stellar populations. In particular, the photons with energies between $11.2-13.6$~eV, known as the Lyman Werner radiation, can travel long distances and dissociate the molecular hydrogen over large scales. The  strength of such flux has been investigated in the numerous studies \citep{2001ApJ...546..635O,2010MNRAS.402.1249S,2014MNRAS.443.1979L}.


It has been shown recently that keeping the collapse completely isothermal with cooling only due to atomic hydrogen  is difficult. This is due to the combined effect of considering realistic spectra emitted from Pop II stars where the H$^-$ photodissociation is disfavoured  \citep{Sugimura14,Agarwal2014} as well as due to the more realistic initial conditions in 3D simulations \citep{Latif201408}, where the degree of ionization is enhanced by the virialization shocks. Hence, it stimulates $\rm H_2$ formation via the H$^-$ route and increases the required strength of the radiation background. Considering both effects, the critical value of the UV flux is now a few times $\rm 10^4$ in units of $\rm J_{21}$ \citep{Latif201408}. This is about two orders of magnitude higher than previous estimates, and the formation sites of direct collapse black holes become exceedingly rare once such thresholds are adopted \citep{2008MNRAS.391.1961D,2012MNRAS.425.2854A,2014MNRAS.442.2036D}. It is therefore important to explore whether direct collapse occurs only under isothermal conditions or if additional processes can also help to build up a massive central object.

Indeed, recent simulations suggest that massive to supermassive stars may form even in the presence of a moderate UV flux. In this case, an initially isothermal collapse in the presence of atomic cooling may still yield a significant mass supply to the center of the halo, from which stars of $10^3-10^4$~M$_\odot$ can efficiently form \citep{Latif2014ApJ}. In such a scenario, large accretion rates of $\rm 0.1-1 ~M_{\odot}/yr$ occur on a spatial scale of 1 pc and the core of the halo is cooled by the molecular hydrogen. In the presence of such rapid accretion, the central part of the disk can potentially form a massive star and fragmentation may occur on larger scales. We expect the formation of a rapidly rotating disk surrounding such stars as a result of angular momentum conservation as proposed by \cite{Oh2002} and \cite{Volonteri2005}.  The properties and the stability of the disk forming in massive primordial halos were first explored by  \cite{Lodato2006}, finding that the mass of the central object depends on halo properties such as mass, spin and its ability to cool. They further suggested that low spin halos are more efficient in forming a massive central object. While the detailed properties of rapidly rotating disks are currently unknown, their thermal and dynamical properties will be highly relevant once the central star becomes massive and dominates the gravitational potential. Even in the presence of moderate accretion rates, viscous heating can raise the gas temperature and help to stabilize the disk against fragmentation \citep{Latif2014Disk}. This effect will be considerably enhanced in the presence of larger accretion rates and more massive central objects. This may have an important implications for the formation of massive central objects.

In this study, we therefore consider an atomic cooling halo exposed to a moderate UV background, implying an initially isothermal collapse with high accretion rates, as expected from numerical simulations \citep{Latif2014ApJ}. Due to the conservation of angular momentum, we assume that a self-gravitating disk forms in the central region, and we expect most of the disk to be initially molecular due to the self-shielding against the radiation background. Using a simple analytical model, we investigate the properties of these disks in the presence of central massive objects, and explore the impact of viscous heating on the thermal evolution. Particularly, we demonstrate that viscous heating can dissociate the molecular hydrogen and leads to a transition towards an atomic cooling regime. Various stages of the disk evolution with a central object of 10, 100 and $\rm 10^4~M_{\odot}$ are investigated. The latter can be considered to represent a time evolution due to the growth of the central object. We study the stability of the disk by computing the viscous parameter $\alpha$. The masses of the clumps forming inside the disk are estimated and their fate is determined by comparing the migration and collapse time scales. 

This article is organized in the following way. In  section~2, we discuss our theoretical framework and its main assumptions. In section~3, we present our main results. In subsection 3.1, we discuss the properties of the disk in the presence of a 10 solar mass star. We further describe the intermediate and late stages of the disk evolution when the central becomes more massive in subsections 3.2 and 3.3. The potential caveats in our model are discussed in section 3.4. In section~4, we summarize the main conclusions and discuss future perspectives.

\section{Theoretical framework}
In massive primordial halos,  a disk is expected to form as a consequence of gravitational collapse due to the conservation of angular momentum \citep{Oh2002,Volonteri2005,Lodato2006}. It is confirmed from three dimensional cosmological simulations that a rotationally supported disk forms in halos of $\rm 10^7-10^8~M_{\odot}$ illuminated by a moderate UV flux emitted by Pop II stars  at z $\sim$ 15 \citep{Regan2014,Latif2015d}. It is found that the rotational support is enhanced in the presence of $\rm H_2$ cooling compared to the atomic line cooling case.  Under these conditions, an initially isothermal collapse occurs but  the central core is cooled by the molecular hydrogen. Mass inflow rates of $\rm \sim 0.1~M_{\odot}/yr$ are observed in the presence of a moderate UV flux and can be maintained for at-least 30,000 years after the initial collapse \citep{Latif2014ApJ,Latif2015d}.

Motivated by these studies, we here employ an analytical model to investigate the properties of a self-gravitating accretion disk  embedded in  large accretion flows \citep{Inayoshi2014,Latif2014Disk}.  The disk is formed in massive primordial halos irradiated by a background UV flux of strengths 100-1000 (in units of $\rm J_{21} = 10^{-21}~erg/cm^2/s/Hz/sr$) at z=15.
The stability of a self-gravitating accretion disk can be measured with the Toomre Q parameter \citep{Toomre1964} given as
\begin{equation}
 Q = \frac{c_{s} \Omega}{\pi G \Sigma}.
\label{eq1}
\end{equation}
Here $\Omega$ and $c_s$ are the orbital frequency and the sound speed of the disk. G is the gravitational constant and $\Sigma$ is the surface density of the disk, which is given as
\begin{equation}
 \Sigma = \frac{\dot{M}_{tot}}{3 \pi \nu}.
\label{eq2}
\end{equation}
$\dot{M}_{\rm tot}$ is the mass accretion rate and $\nu$ is the viscosity in the disk resulting from the gravitational stresses. We presume that the disk is marginally stable with Toomre Q = 1 and its scale-height can be estimated assuming that the disk is in hydrostatic equilibrium in the vertical direction via
\begin{equation}
 H = \frac{c_s}{\Omega} .
\label{eq3}
\end{equation}

The temperature of the disk is computed by assuming that the viscous heating rate is approximately equal to the disk cooling rate. The viscous heating rate for a self-gravitating disk can be estimated by solving Eq.~(50) given in \cite{Lodato2007}. They provide the heating rate per surface area given as:
\begin{equation}
Q_+ = \nu \Sigma (R \Omega')^2.
\label{eq5}
\end{equation}
In the case of Keplarian rotation, $\Omega_K = \sqrt{\frac{GM_{*}}{R^3}}$. As realistic disks are at least partially supported also via turbulent and thermal pressure, we assume here that the rotation is not entirely Keplerian, but adopt an efficiency parameter $\epsilon_K=\Omega/\Omega_K\sim0.5$ which describes the deviations of the angular velocity $\Omega$ from the Keplerian rotation. Using such a rotational profile, the viscous heating rate can be evaluated as
\begin{equation}
Q_+ = \frac{9}{4} \nu \Sigma \Omega^2.
\label{eq6}
\end{equation}
It may also be noted that similar expressions for the viscous heating rate are given by \cite{Pringle1981} and \cite{Balbus1998}.\footnote{We note that there is a factor of two difference between the expressions given by \cite{Pringle1981} and \cite{Lodato2007} which comes from the fact whether one considers one side of the disk or both sides. We do not distinguish here between both sides of the disk and therefore use the expression given by \cite{Lodato2007}.} The cooling rate per surface area of the disk can be computed as follows: 
\begin{equation}
 Q_- = 2 H \Lambda_{\rm H/H_{2}}.
\label{eq61}
\end{equation}
Here $\Lambda_{\rm H/H_{2}}$ is the atomic or molecular cooling rate  in units of $\rm erg/cm^3/s$ and $H$ is the height of the disk. To compute the viscosity of the disk, we adopt the standard formalism  given by \cite{Shakura73} as
\begin{equation}
 \nu = \alpha c_s H,  
\label{eq7}
\end{equation} 
where $\alpha$ is the viscous parameter and is calculated by combining equations \ref{eq1}, \ref{eq2}, \ref{eq3} and \ref{eq7} as
\begin{equation}
\alpha = \frac{\dot{M}_{tot} G}{ 3 c_s^3}. 
\label{eq9}
\end{equation}

Solving the thermal balance equation (i.e., $\rm Q_+ = Q_-$) with respect to $\Omega$ in the case of molecular cooling, we get the following expression: 
\begin{equation}
\Omega^2 = \frac{24 \pi c_s}{9 \dot{M}_{tot}} \frac{(8 \times 10^9)^{0.45}} {\sqrt{4 \pi m_p G}} \Lambda_{\rm H_2}  \label{H2}
\label{eq8}
\end{equation} 
For the densities given in the disk ($\rm > 10^8 cm^{-3}$), we use the optically thick $\rm H_2$ line cooling by employing the opacity factor $min \left[1,\left(n_{H_2}/(8 \times 10^9~cm^{-3})\right)^{-0.45}\right]$ given by \cite{2004MNRAS.348.1019R}. We use the high density limit for $\rm H_2$ cooling given by \cite{1998A&A...335..403G} adopted from \cite{Hollenbach79}. We further  assume that the transition from the molecular to the atomic cooling regime occurs once the gas temperature reaches a value of $\rm \geq 4000$~K. For atomic cooling regime, solving the thermal balance equation yields the following relation:
\begin{equation}
\Omega^2 = \frac{3 \dot{M}_{tot}(2 \pi G m_p)^{2.5}}{8 \pi c_s} f(T) \Lambda_{H}  \label{atomic}
\label{eq81}
\end{equation}
where $f(T)$ is given as \citep{2001ApJ...546..635O, Inayoshi2014}
\begin{equation}
f(T)= 5.0 \times 10^{-11} T^{1.2}exp\left(\frac{-1.27 \times 10^{5}}{2 T}\right).
\end{equation}
Here $\Lambda_{H}$ is the atomic cooling function. We use the expression given in the appendix B of \cite{1997NewA....2..209A} originally computed by \cite{Cen92} to take into account the cooling due to the collisional excitation of hydrogen atoms (known as atomic line cooling). We have checked using a one-zone model that the latter indeed yields the dominant contribution in this regime \citep{2014MNRAS.439.2386G}. We note that the differences between Eqs.~(\ref{H2}) and (\ref{atomic}) arise due to the multiplication with the species abundances, as well as the correction factor due to the optical depth. As we are considering a case with a large inflow rate onto the disk, the central star will form and grow via rapid accretion, and the central star will also dominate the gravitational potential at least on the scales considered here. The surface density of the disk can be computed as
\begin{equation}
 \Sigma = \frac{c_s}{\pi G}  \sqrt{\frac{GM_{*}}{R^3}} ,
\label{eq20}
\end{equation} 
and density inside the disk is
\begin{equation}
n = \frac{\Sigma}{4 \pi m_p c_s}  \sqrt{\frac{GM_{*}}{R^3}}.
\label{eq21}
\end{equation}
We estimate the initial masses of the clumps as
\begin{equation}
M_{c,i} = \Sigma H^2.  
\label{eq13}
\end{equation} 
These clumps can grow via accretion and we compute the maximum clump mass, the so-called the gap-opening mass, by using the relation given in \cite{2007MNRAS.374..515L}:
\begin{equation}
M_{c} =  M_{c,i} \left[12 \pi \frac{\alpha}{0.3}\right]^{1/2} \left(\frac{R}{H}\right)^{1/2}.
\label{eq14}
\end{equation} 
The mass accretion rate onto the clumps ($\dot{M}_c$) is estimated as:
\begin{equation}
 \dot{M}_c =  \frac{3}{2} \Sigma \Omega \left( R_H \right)^2,
\label{eq16}
\end{equation}
where R$_H$ is the Hill radius given as
\begin{equation}
 R_H =  R \left( M_c /30 M_* \right)^{1/3}.
\label{eq17}
\end{equation}
$M_*$ is the mass of the central star. The Kelvin-Helmholtz (KH) contraction time scale is estimated as :
\begin{equation}
t_{KH} =  \frac{M_c}{\dot{M}_c}  \sim  10^4 \left(\frac{M_c}{30 M_{\odot}}\right) \left[\frac{\dot{M}_c}{3\times 10^{-3} M_{\odot} yr^{-1}}\right]^{-1}~{\rm yr}.
\label{eq151}
\end{equation}
The Kelvin-Helmholtz (KH) contraction time scale given above is equivalent to the mass accretion time scale and is valid in the presence of high accretion rates ($\rm 0.1~M_{\odot}/yr$), which remain in the stage of adiabatic evolution as long as the accretion rate is high. In the case of accretion rates below $10^{-2}$~M$_\odot$/yr, we compute the KH timescale using the following expression \citep{Hosokawa12}: 
\begin{equation}
t_{KH} = \frac{G M_*^2}{R_* L_*}, 
\label{eq15}
\end{equation}
where  $M_*$ is the mass, $R_*$ is the radius and $L_*$ is the luminosity of the protostar. We use Eq.~(4) of \cite{Hosokawa12} to compute the luminosity and their Fig. 5 to estimate the radius of the star. The clumps formed inside the disk will migrate inward provided that the migration time scale is shorter than the time for the clumps to contract and form a protostar. The migration of clumps forming in self-gravitating disks has been investigated in detail in the context of planet formation as a result of gravitational instabilities \citep{2002ApJ...572..566R,2009ApJ...704..281R}. In the theory of planet-disk interactions, the idealized migration time scales are well known. If the mass of the planet is small and the perturbations induced by the clumps are in the linear regime, one speaks of type I migration. On the other hand, if the planet becomes massive enough to open a gap in the disk, it migrates inward roughly on the viscous time scale. This is called the type II migration and its timescale can be estimated as \citep{Lin86}
\begin{equation}
t_{mig} =  \frac{1}{3 \pi \alpha} \left( \frac{R} {H} \right)^{2} \frac{2 \pi}{\Omega}.
\label{eq18}
\end{equation}
The migration time scales are usually derived assuming that the planet acts like a point mass in the disk and the density profile in the disk is smooth. Nonetheless, the results from the numerical simulations are in a good agreement with these estimates \citep{Baruteau11,Zhu2012}. We use the expression given in equation \ref{eq18} to compute the clump migration time scale. Our choice of the migration time scale is rather conservative as the time for the type II migration is slower than the type I migrations. We note that this timescale may be even shorter for clumpy disks, therefore potentially enhancing the importance of migration.

To quantify the impact of tidal forces by the central star, we compute the Roche limit given as
\begin{equation}
R_{Roche} =  1.26 \times R_c \left( M_* / M_c \right)^{1/3},
\label{eq19}
\end{equation}
where $R_{\rm c}$, $M_{\rm c}$ are the radius and the mass of the secondary clumps and $M_{\rm *}$ is the mass of the central star. 

We estimate the photo-evaporation of the disk by the central star in the following way. For high accretion rates (i.e. $\rm \geq 0.01~M_{\odot}/yr$), the stellar radius  almost linearly increases with mass and consequently these stars have large radii with low surface temperatures of about 5000 K \citep{Schleicher13,Hosokawa13}. They  emit close to  the Eddington luminosity given as \citep{Ferrara14}
\begin{equation}
L_{Edd} =  1.5 \times 10^{38} \left( M_* / M_{\odot} \right)~ erg/s.
\label{eq201}
\end{equation}
To obtain an upper limit on the number of ionizing photons per second, we assume here that all photons are emitted at a frequency of $13.6$~eV.   The mass evaporation rate of the disk is computed using the expression  given by \cite{Hollenbach94}:
\begin{equation}
\dot{M}_{PE} = 1.3 \times 10^{-5} \left(\frac{\phi_{UV}}{10^{49}~s^{-1}} \right)^{1/2}  \left(\frac{M_{*}}{10~M_{\odot}}\right)^{1/2}~M_{\odot}/yr 
\label{eq211}
\end{equation}
Dividing the disk mass by the photo-evaporation rate, we estimate the disk photo-evaporation time as
\begin{equation}
t_{PE} =  \frac{M_{disk}}{\dot{M}_{PE}} ~yr.
\label{eq22}
\end{equation}
The expression given here is only a lower limit, due to the assumption of peak emission at 13.6 eV. As rapidly accreting supermassive stars are expected to have cool atmospheres with a temperature of 5000~K, the UV emission may be considerably suppressed, reducing the potential effect of photoevaporation \citep{Schleicher13,Hosokawa13}. We expect the photo-evaporation time of individual gravitationally bound clumps to be even longer, due to their enhanced surface and column densities.




\section{Results}
We present here the results from our model described in the previous section. The properties of the disk forming inside a massive primordial halo of $\rm 10^7-10^8~M_{\odot}$ illuminated by a moderate Lyman Werner flux are investigated here. We presume that the gas inflow rate to the disk is $\rm 0.1~M_{\odot}/yr$ (unless stated otherwise), as found in numerical simulations investigating such conditions \citep{Latif2014ApJ}. In the presence of such accretion rates, the central star is expected to be on the Hayashi track, implying a cool atmosphere with $\sim5000$~K where radiative feedback is inefficient \citep{Hosokawa12, Schleicher13}. The gas in the disk is of a primordial composition and we consider the gas to be fully molecular in the initial stage. We examine the disk properties over a range of evolutionary stages considering a central star of 10, 100 \& $\rm 10^4~M_{\odot}$. Particularly, we explore how the viscous heating in the presence of the central star influences the thermal properties of the disk. We note that a very similar evolution is expected if the central object is not a protostar, but already a massive black hole. In this case, the viscous heating would be the same, while the self-shielding of the disk would prevent a strong impact of the radiation onto the disk. The results for the various cases are discussed in the following subsections.

\subsection{Early stage of disk evolution with $\rm 10~M_{\odot}$ central star}
\begin{figure*}
\hspace{-6.0cm}
\centering
\begin{tabular}{c c}
\begin{minipage}{6cm}
\vspace{0.2cm}
\includegraphics[scale=0.7]{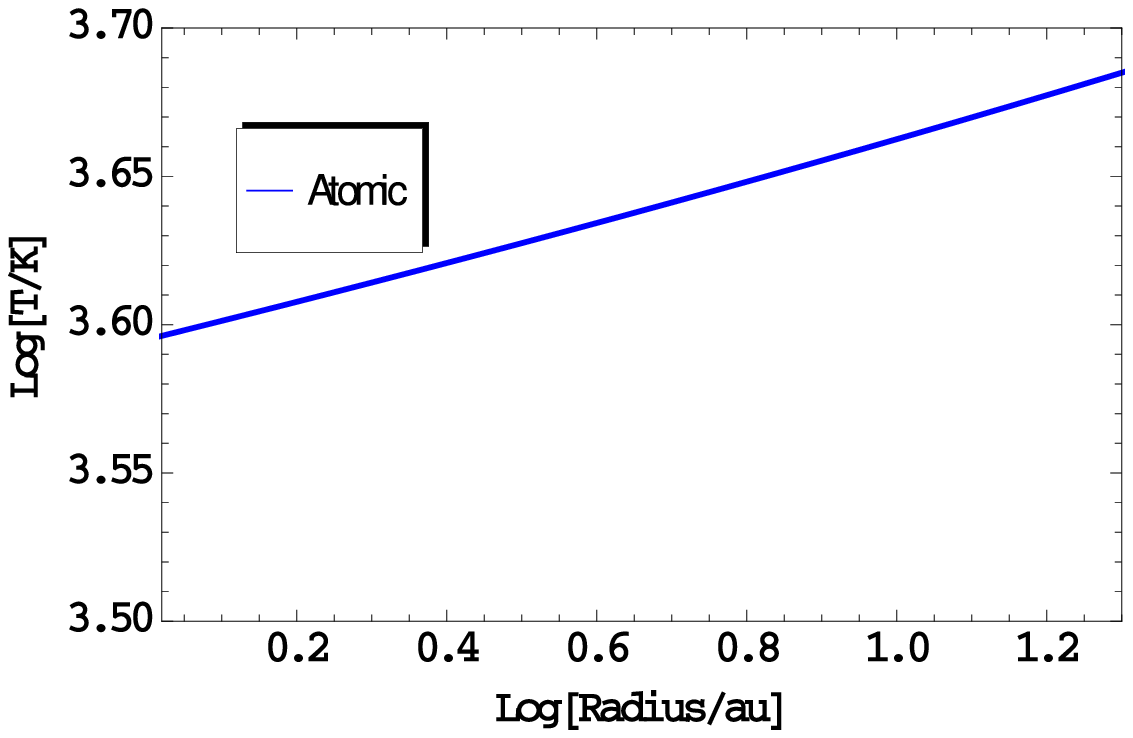}
\end{minipage}&
\begin{minipage}{6cm}
\hspace{2cm}
\includegraphics[scale=0.7]{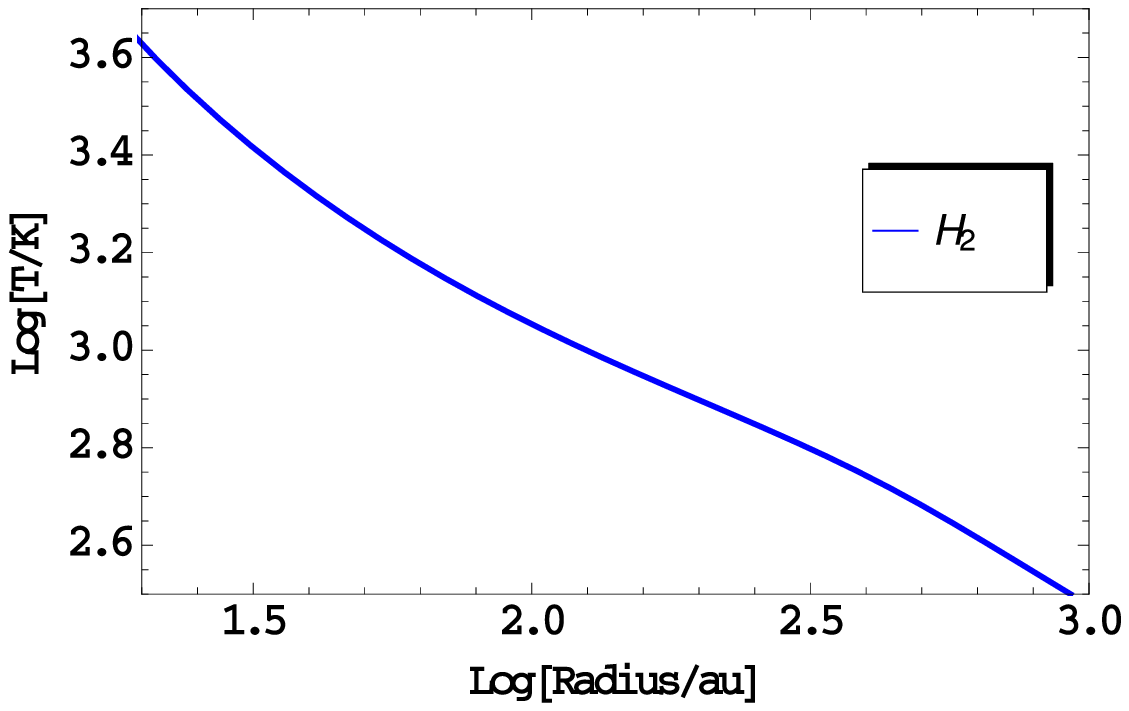}
\end{minipage} \\
\begin{minipage}{6cm}
 \vspace{-0.14cm}
\includegraphics[scale=0.7]{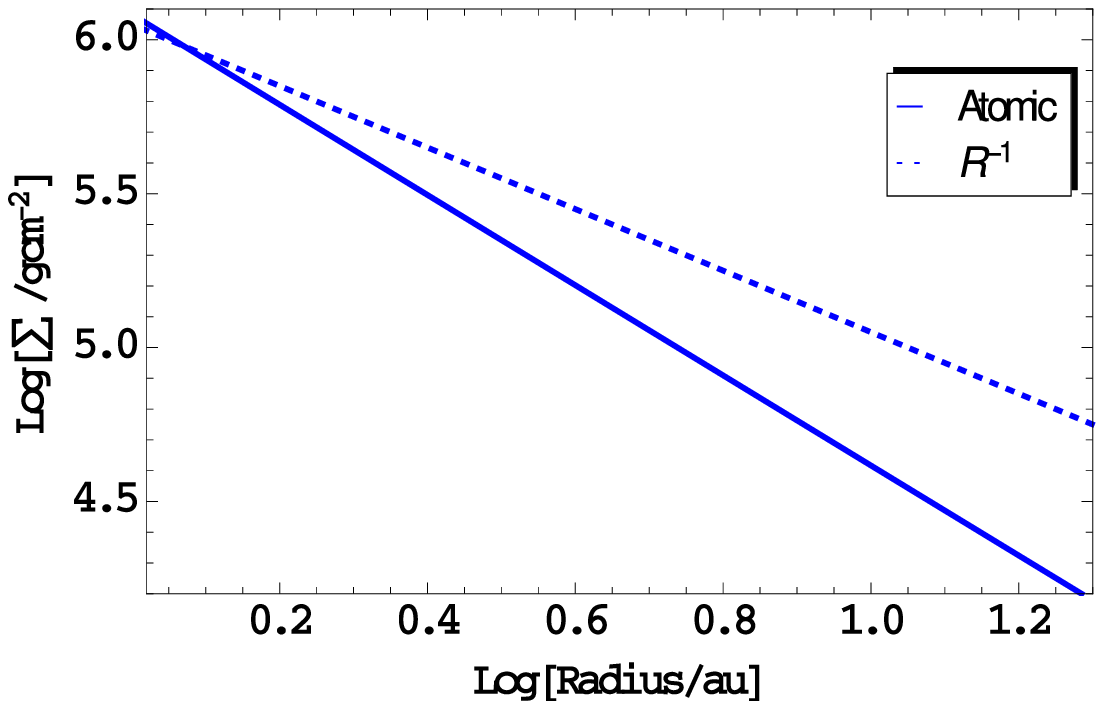}
\end{minipage}&
\begin{minipage}{6cm}
 \hspace{2cm}
\includegraphics[scale=0.7]{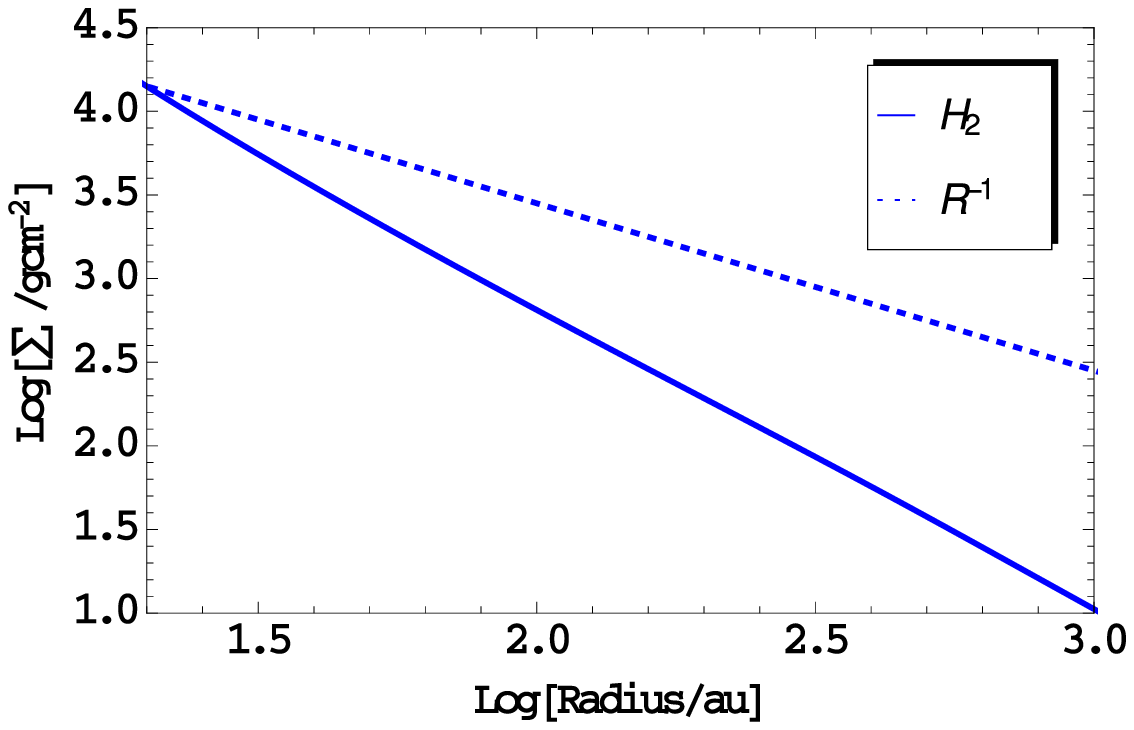}
\end{minipage} \\
\begin{minipage}{6cm}
\vspace{-0.1cm}
\includegraphics[scale=0.7]{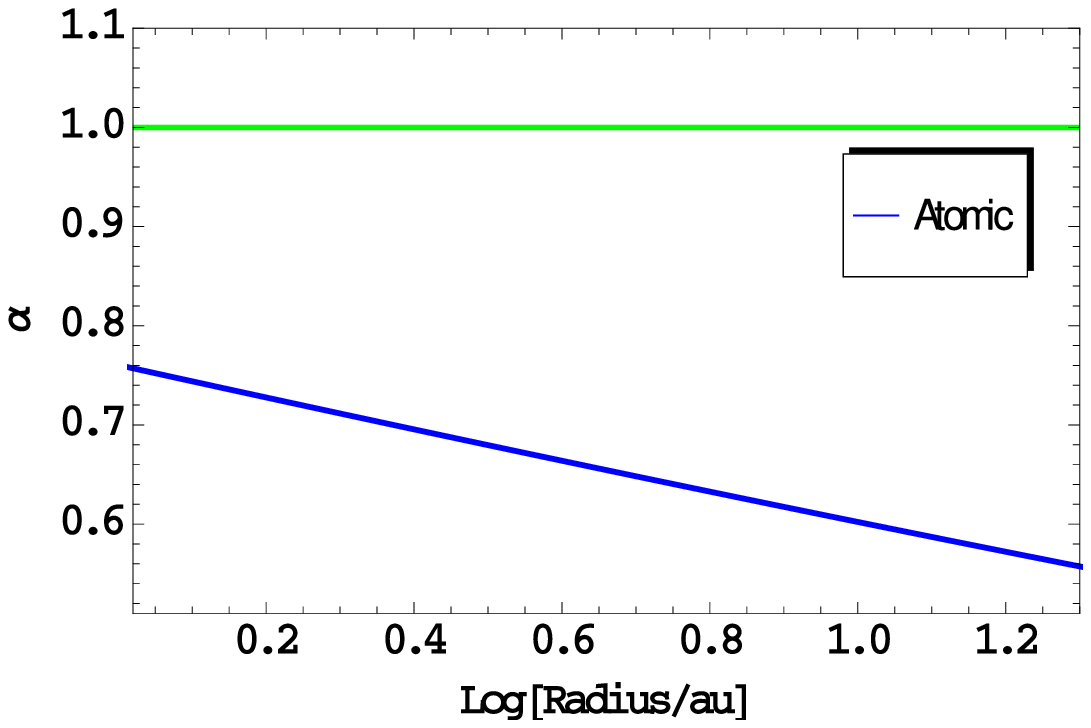}
\end{minipage}&
\begin{minipage}{6cm}
\hspace{2cm}
\includegraphics[scale=0.7]{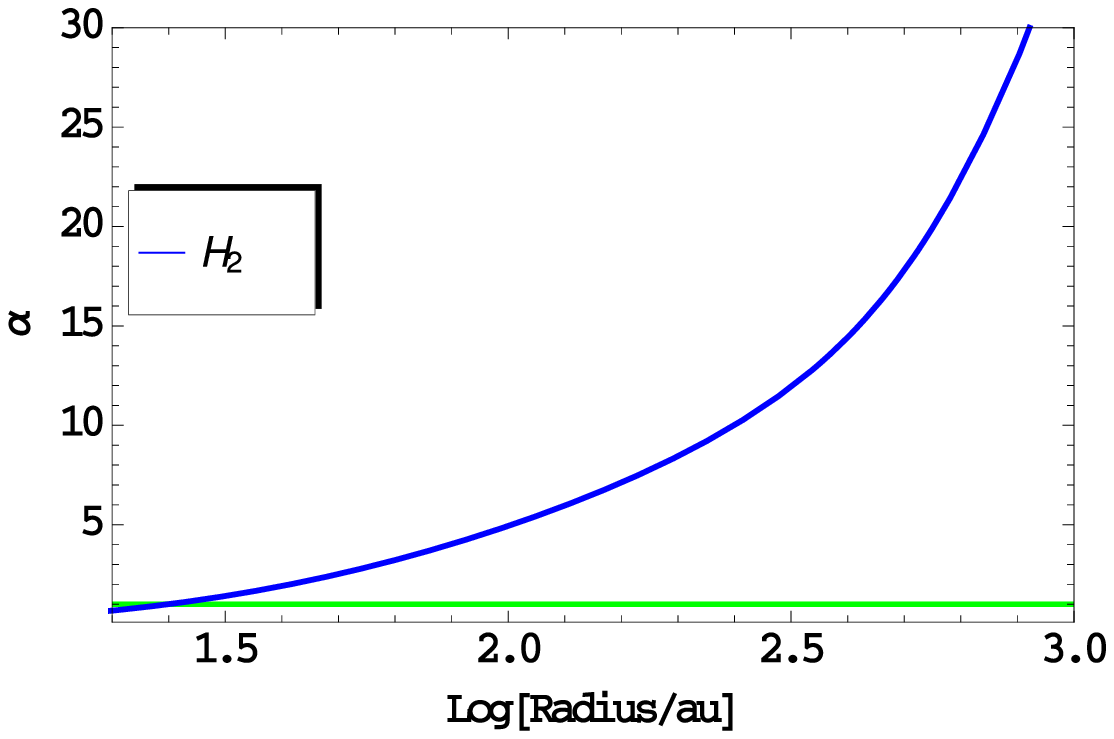}
\end{minipage} \\
\begin{minipage}{6cm}
\vspace{0.2cm}
\includegraphics[scale=0.7]{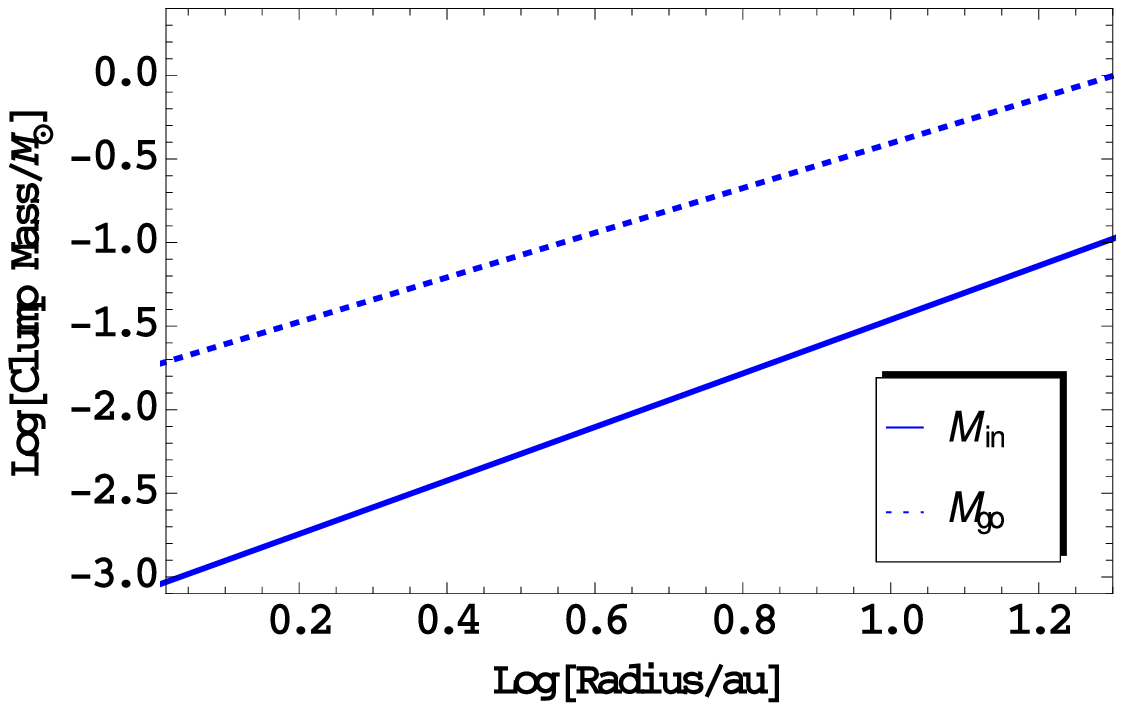}
\end{minipage}&
\begin{minipage}{6cm}
\hspace{2cm}
\includegraphics[scale=0.7]{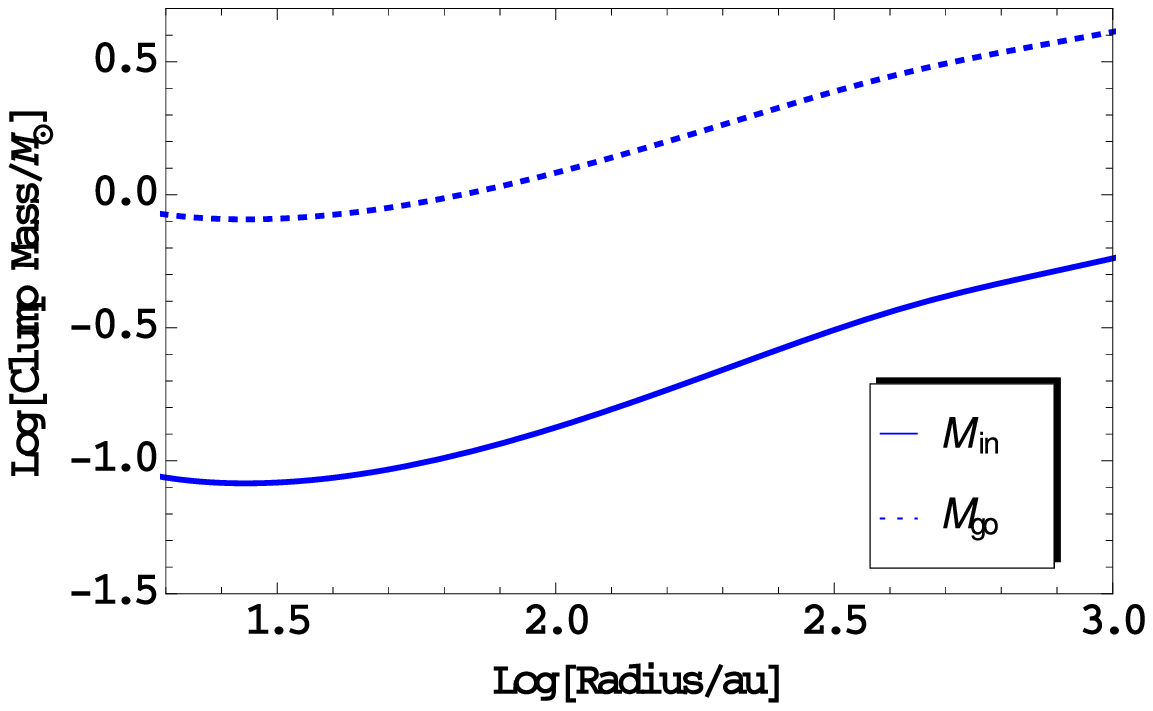}
\end{minipage} 
\end{tabular}
\caption{The disk properties for the central star of 10 $\rm ~M_{\odot}$. The temperature, the surface density, the viscous parameter $\alpha$ and the clump masses are shown here. The left panel shows the atomic while the right panel depicts the $\rm H_2$ cooling regime.}
\label{fig0}
\end{figure*}

The thermal properties of the disk with a central star of $\rm 10~M_{\odot}$ are shown in figure \ref{fig0}. It is found that above $20$~AU the disk is cooled by molecular hydrogen and the temperature in the disk varies from about 300 to 4500 K. At this point, the viscous heating becomes dominant, molecular hydrogen gets collisionally dissociated and is not able to cool the gas anymore. At radii below 20 AU, atomic line cooling starts to become effective and keeps the gas temperature above 4000 K. It shows that in the presence of high accretion rates, the molecular hydrogen is collisionally dissociated and a transition from the molecular to the atomic line cooling takes place. The surface density in the disk varies from 1 to  $\rm 10^6~g/cm^2$.  For comparison, we show a profile with $\rm R^{-1}$ as expected for a Mestel disk \citep{Mestel63}. The profile for the surface density is steeper than for the Mestel disk due to the non-isothermal evolution.

To assess the stability of the disk, we compute the viscous parameter $\alpha$ for both the  molecular and the atomic cooling regime, which is shown in figure \ref{fig0}. Our choice for the critical value of  $\alpha$ is 1 (see discussion in subsection~\ref{uncertainties}), which is motivated by numerical simulations including a detailed model for thermal processes showing that disks fragment at $\alpha = 1$ \citep{Zhu2012},  see also \cite{Gammie01}. The results from our model show that the disk is stable within $20$~AU, hereafter called the stable radius ($\alpha \leq 1$), and becomes unstable at larger radii ($\alpha > 1$). Our estimates for the clump masses in the unstable regime show that clumps with masses between $\rm 0.1-1~M_{\odot}$ are expected to form. We further estimate the gap opening mass (i.e. the mass obtained by the clumps after opening a gap in the disk) and find that it can reach up to a few solar masses. The mass accretion rates onto the clumps range from $\rm 0.01~M_{\odot}/yr$ in the interior of the disk to $ \rm 0.001~M_{\odot}/yr$ in its outskirts. This can be understood from the fact that the gas density in the inner part of the disk is higher than at the outer radii which leads to higher accretion rates. 

\begin{figure*}
\hspace{-6.0cm}
\centering
\begin{tabular}{c c}
\begin{minipage}{6cm}
\includegraphics[scale=0.7]{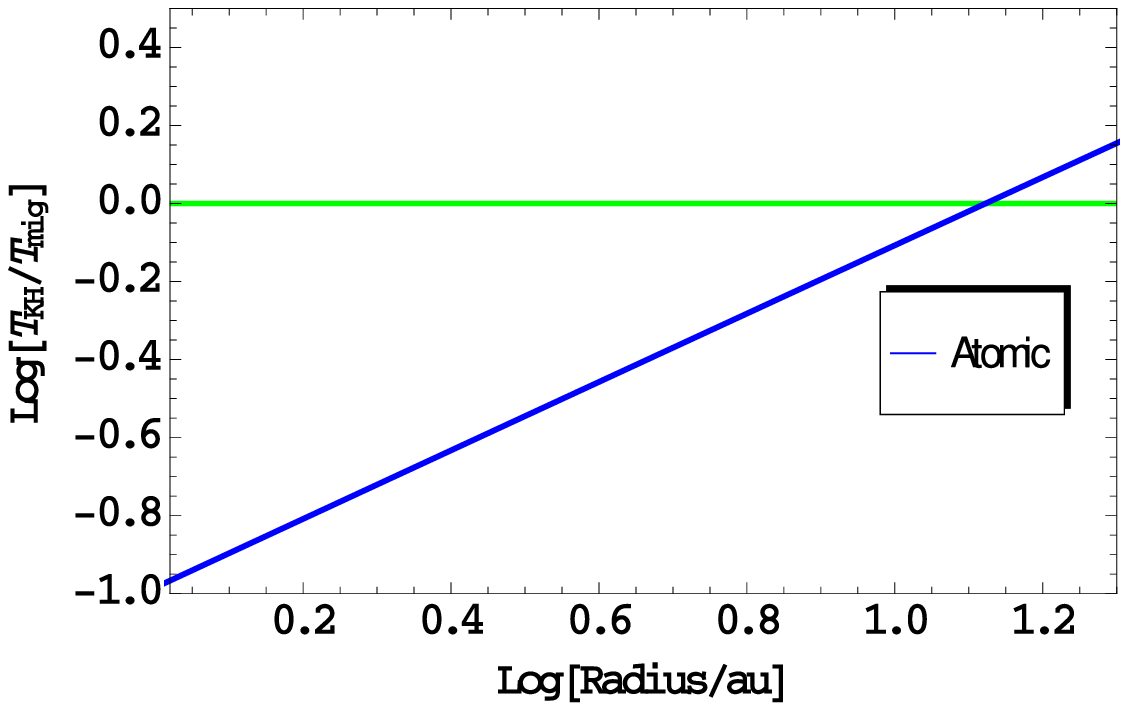}
\end{minipage}&
\begin{minipage}{6cm}
\hspace{2cm}
\includegraphics[scale=0.7]{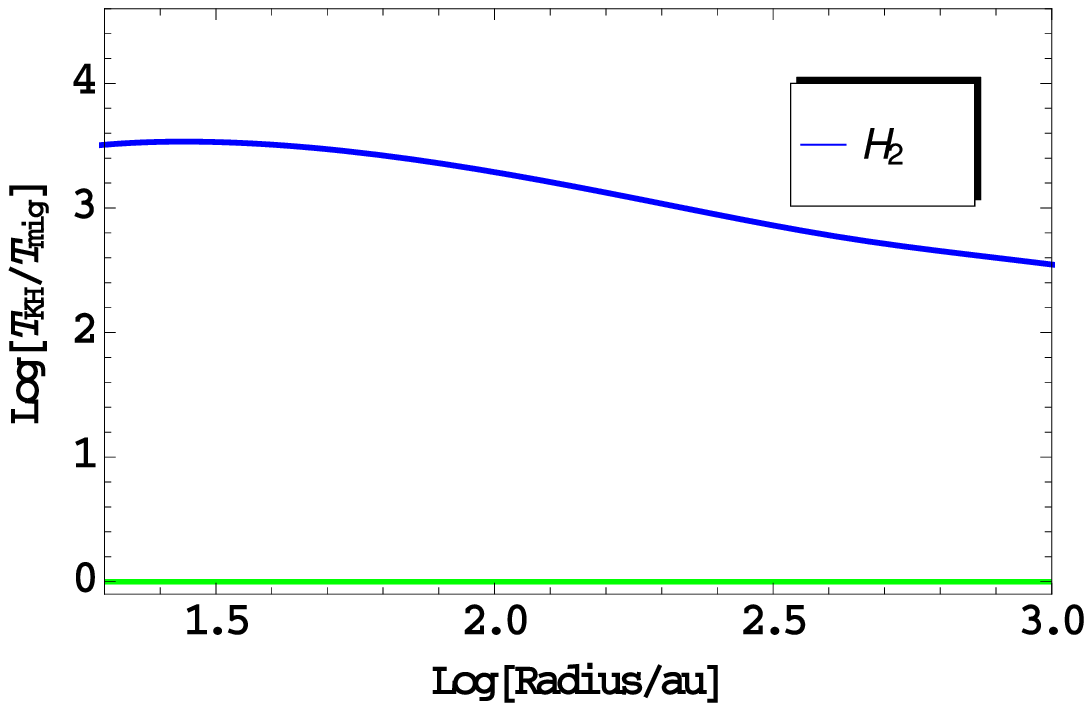}
\end{minipage} \\
\begin{minipage}{6cm}
\vspace{0.2cm}
\includegraphics[scale=0.7]{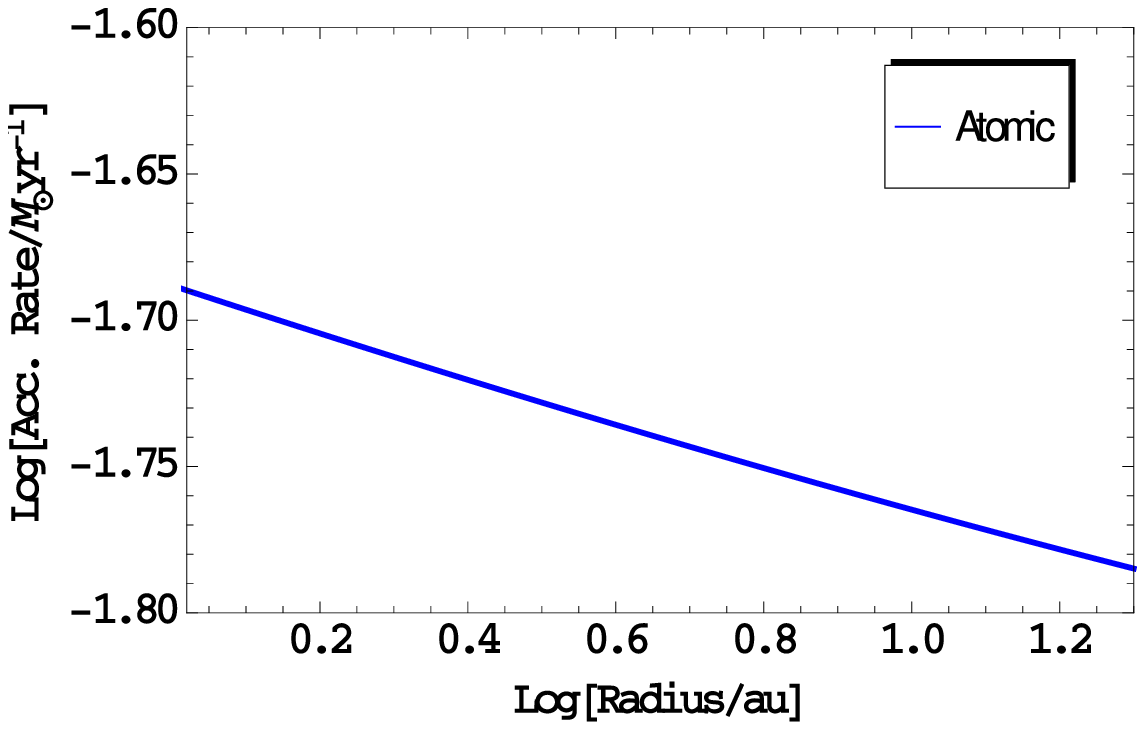}
\end{minipage}&
\begin{minipage}{6cm}
 \hspace{2cm}
\includegraphics[scale=0.7]{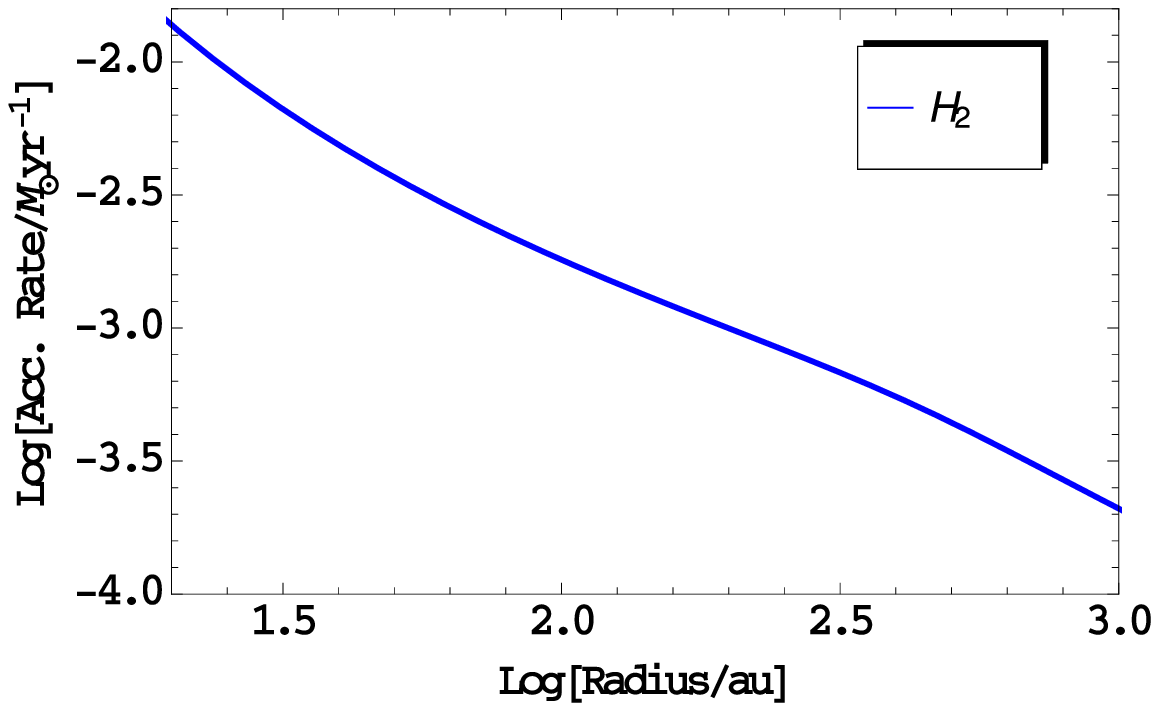}
\end{minipage} \\
\begin{minipage}{6cm}
\vspace{0.2cm}
\includegraphics[scale=0.7]{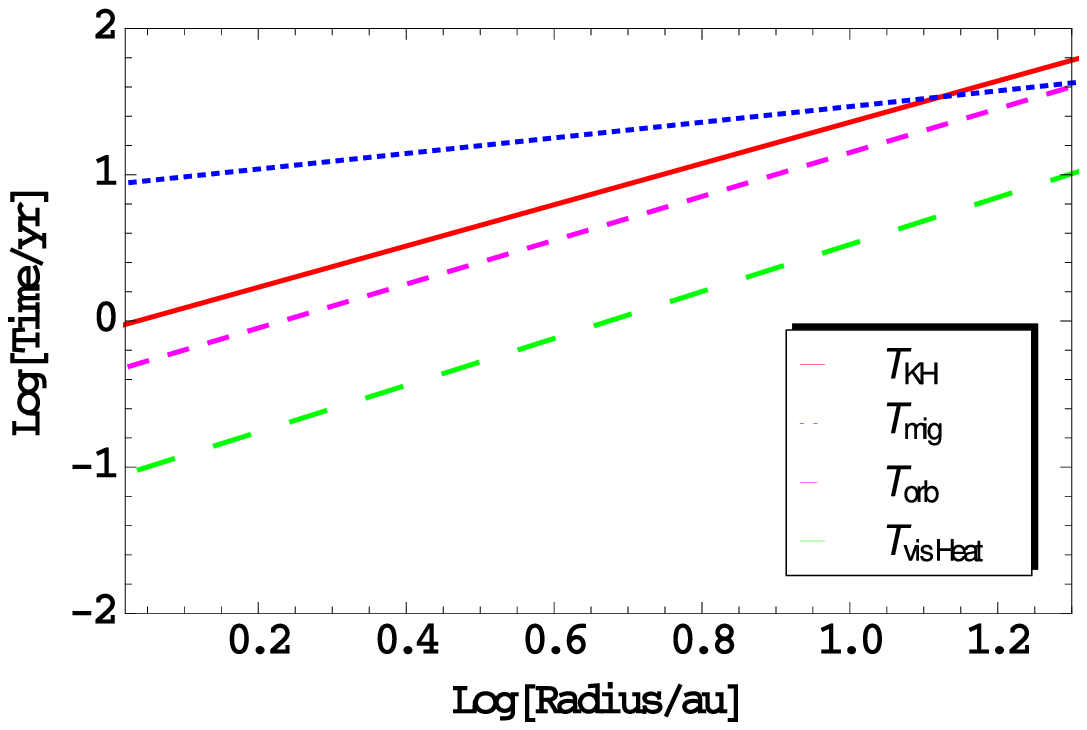}
\end{minipage}&
\begin{minipage}{6cm}
\hspace{2cm}
\includegraphics[scale=0.7]{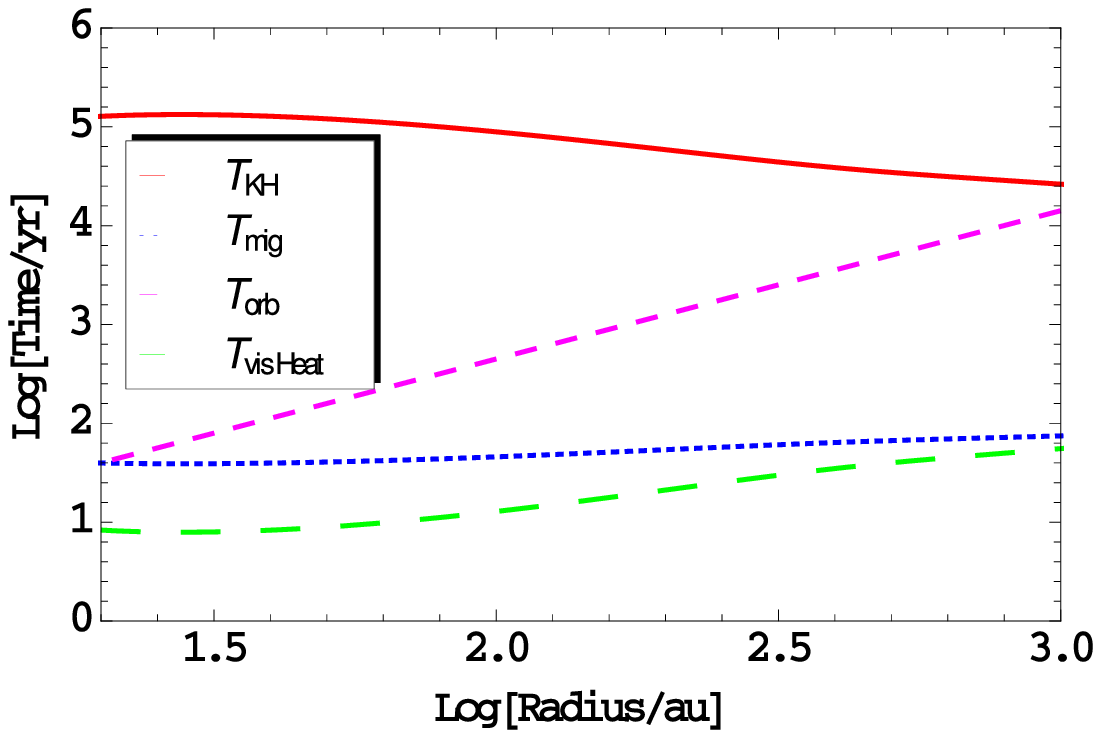}
\end{minipage} 
\end{tabular}
\caption{The disk properties for the central star of 10 $\rm ~M_{\odot}$. The migration, the Kelvin-Helmholtz, the orbital and the viscous time scales as well as the accretion rates onto the clumps are shown here. The left panel shows the atomic while the right panel depicts the $\rm H_2$ cooling regime.}
\label{fig1}
\end{figure*}

To determine whether clumps will be able to survive or migrate inward, we have estimated the migration ($\rm T_{mig}$) and the Kelvin Helmholtz ($\rm T_{KH}$) time scales. The ratio of $\rm T_{KH}$ to $\rm T_{mig}$ is shown in figure \ref{fig1}. We find that this ratio remains larger than one in the unstable region. Therefore, clumps are expected to migrate inward on the viscous time scale. For comparison with other timescales in the problem, we have also computed the viscous heating time, the orbital time and the photo-evaporation time, and compared them with $\rm T_{mig}$ \& $\rm T_{KH}$. We note that the Kelvin Helmholtz time for the clumps remains larger than all other time scales in the $\rm H_2$ cooling regime because of the lower accretion rates and gets inverted within the 10 AU due to the higher accretion rates. In the atomic cooling regime, $\rm T_{KH}$ seems to drop in the interior. However, due to the rather low clump masses in this regime, we expect no significant impact on the further accretion. Overall, we show that the orbital time scale is comparable to $\rm T_{KH}$ in the atomic cooling regime and scales with radius in the $\rm H_2$ cooling regime. It is also noted that the viscous heating time scale is shorter than orbital time scale which means that significant heating can occur within one orbital period. The photo-evaporation time scale of the disk is of the order of $\rm 10^5$ years, much longer than the migration and the viscous time scales.  It is comparable to the Kelvin Helmholtz time scale in the outer parts of the disk. Therefore,  we expect that the impact of photo-evaporation will be negligible during the timescales considered here.

\subsection{Intermediate stage  with $\rm 100~ M_{\odot}$ central star}

Due to the short migration time and the stability in the inner part of the disk, the central star is expected to rapidly grow in mass. We have computed the state of the disk for a central star of a 100 $\rm M_{\odot}$. The properties of the disk are very similar to the case with a central star of $\rm 10~M_{\odot}$, in the presence of an accretion rate of $\rm 0.1~M_{\odot}/yr$, initially the disk is cooled by the molecular hydrogen. Due to the strong viscous heating, $\rm H_2$ cooling becomes inefficient and eventually $\rm H_2$ gets collisionally dissociated. It is found that in this case, the range for atomic cooling increases up to a scale of $40$~AU which is about twice the size of the case with $\rm M_*= 10~M_{\odot}$. In general, the disk is hotter and denser in comparison with the $\rm 10~M_{\odot}$ central star. The disk surface density is about a factor of three higher and reaches a few times $\rm 10^5~g/cm^2$ at 4 AU. The stable radius is further shifted outwards and extends up to $63$~AU. No large differences are observed in the clump masses. The mass accretion rates onto the clumps are about an order of magnitude higher and range from about $\rm 0.02- 0.1~M_{\odot}/yr$.

The comparison of $\rm T_{mig}$ and $\rm T_{KH}$ shows that the clumps forming beyond the stable radius are expected to migrate inward. The Kelvin-Helmholtz time is shorter compared to the previous case because of the higher accretion accretion rates onto the clumps. The viscous heating time is again much shorter compared to all other time scales. It is about  one year and further declines in the disk interior. The photo-evaporation time scale for the disk is about 10,000 years. It is longer than all other time scales and thus its effect on the mass accretion of the central star is considered to be  minor.

\subsection{Late stage with $\rm 10^4 ~ M_{\odot}$ central star}
\begin{figure*}
\hspace{-6.0cm}
\centering
\begin{tabular}{c c}
\begin{minipage}{6cm}
\vspace{0.2cm}
\includegraphics[scale=0.7]{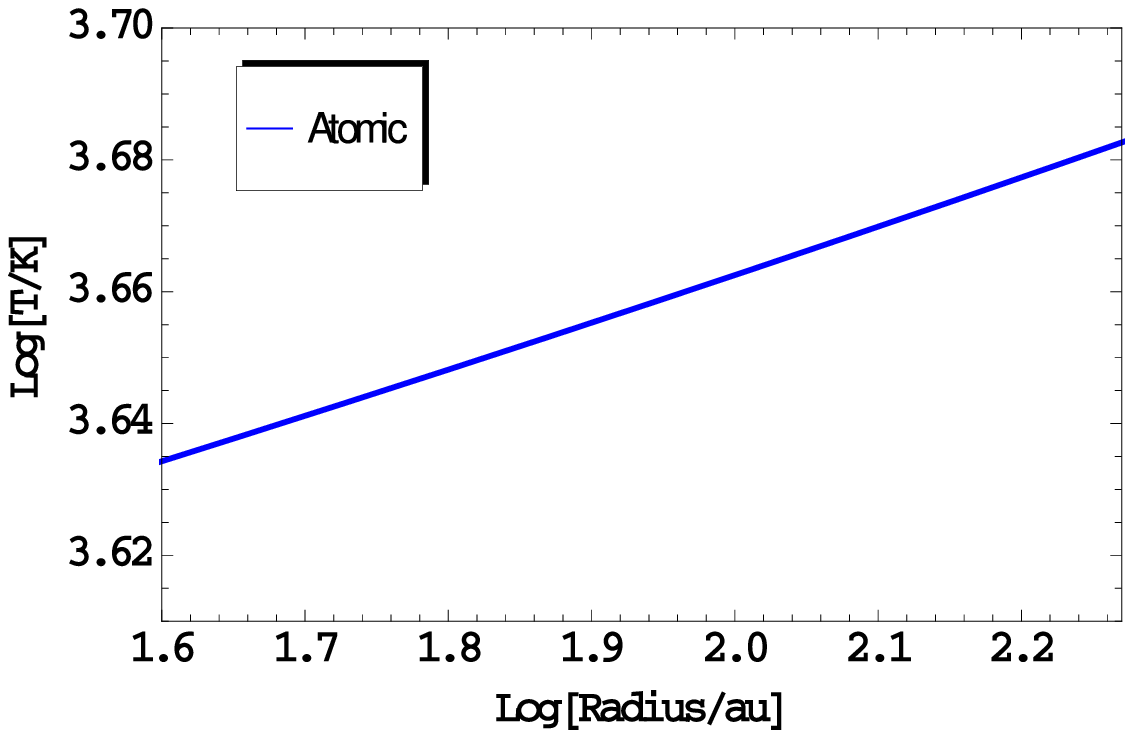}
\end{minipage}&
\begin{minipage}{6cm}
\hspace{2cm}
\includegraphics[scale=0.7]{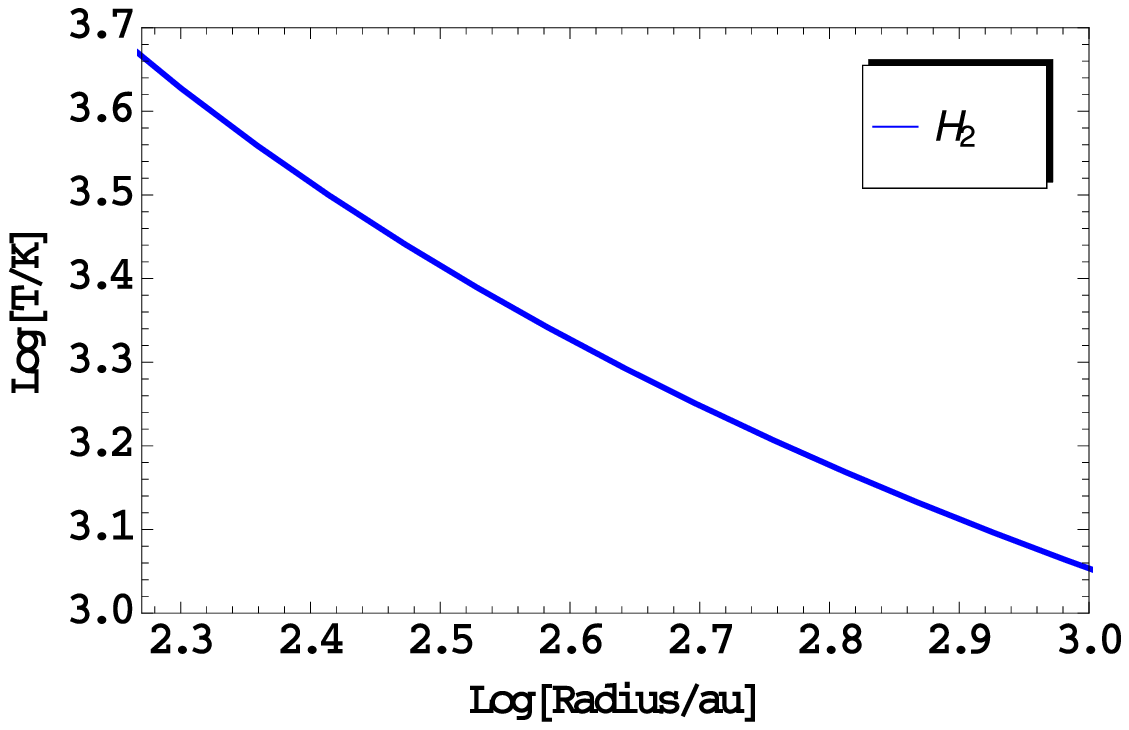}
\end{minipage} \\
\begin{minipage}{6cm}
\vspace{-0.1cm}
\includegraphics[scale=0.7]{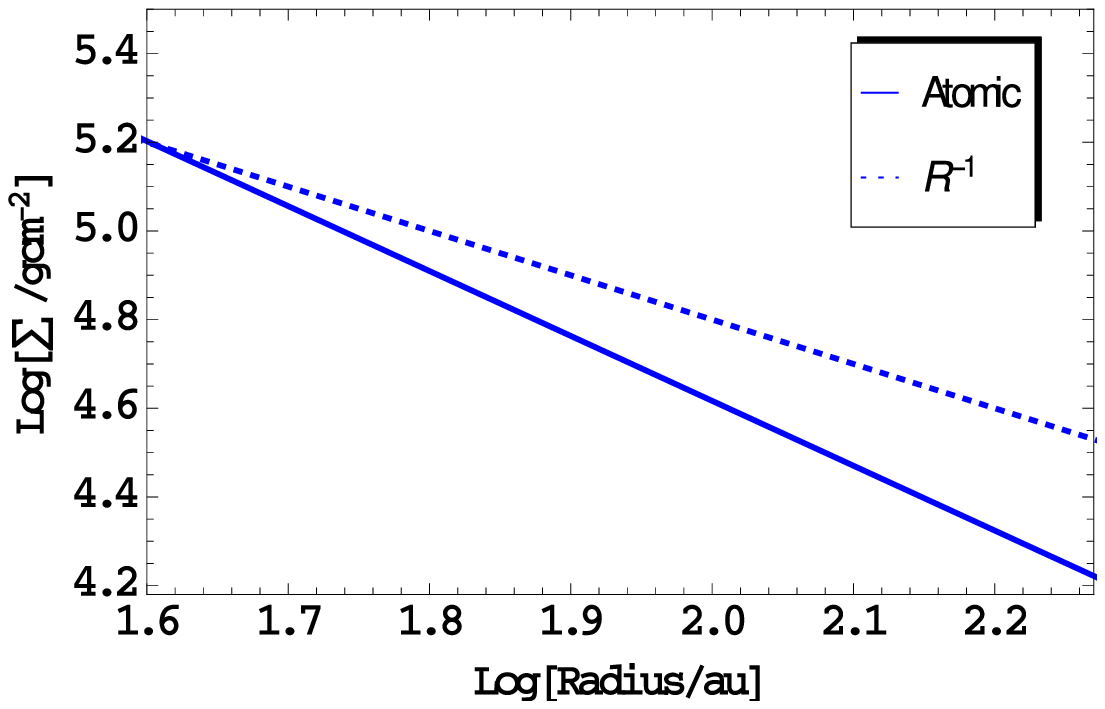}
\end{minipage}&
\begin{minipage}{6cm}
 \hspace{2cm}
\includegraphics[scale=0.7]{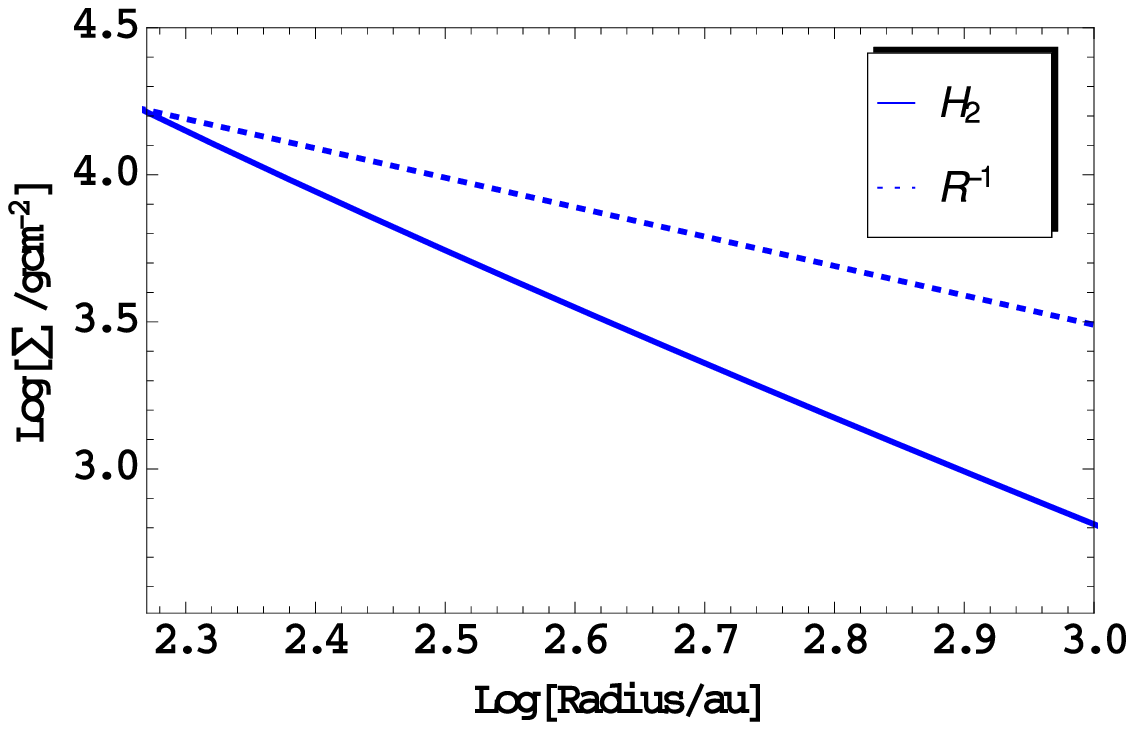}
\end{minipage} \\
\begin{minipage}{6cm}
 \vspace{-0.1cm}
\includegraphics[scale=0.7]{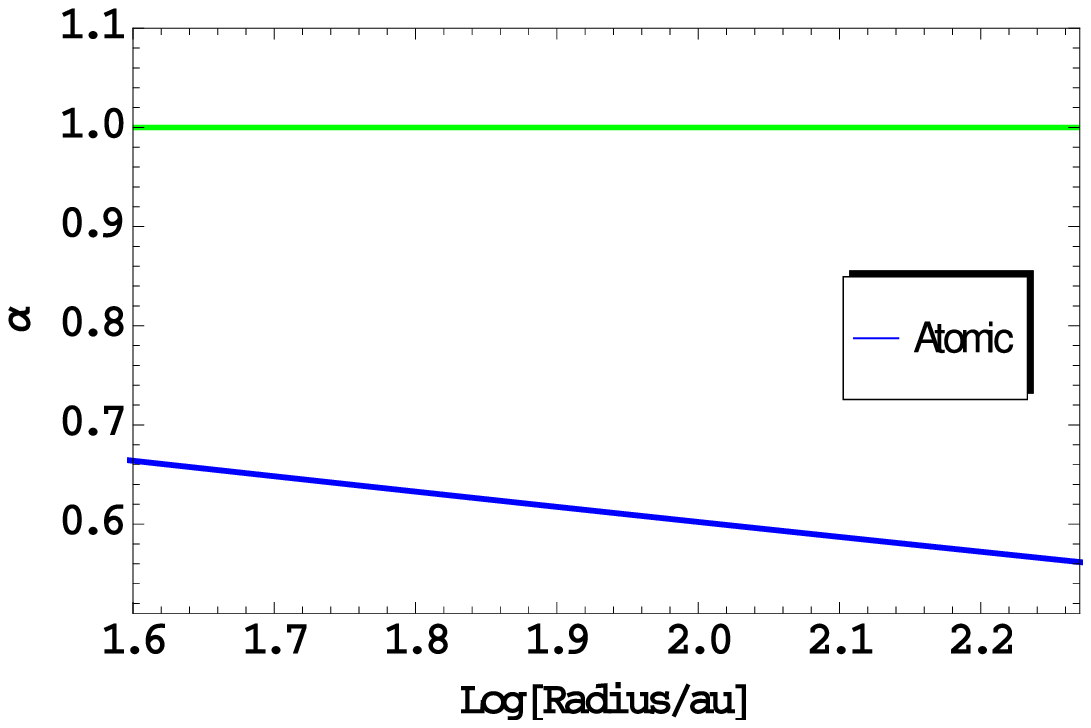}
\end{minipage}&
\begin{minipage}{6cm}
\hspace{2cm}
\includegraphics[scale=0.7]{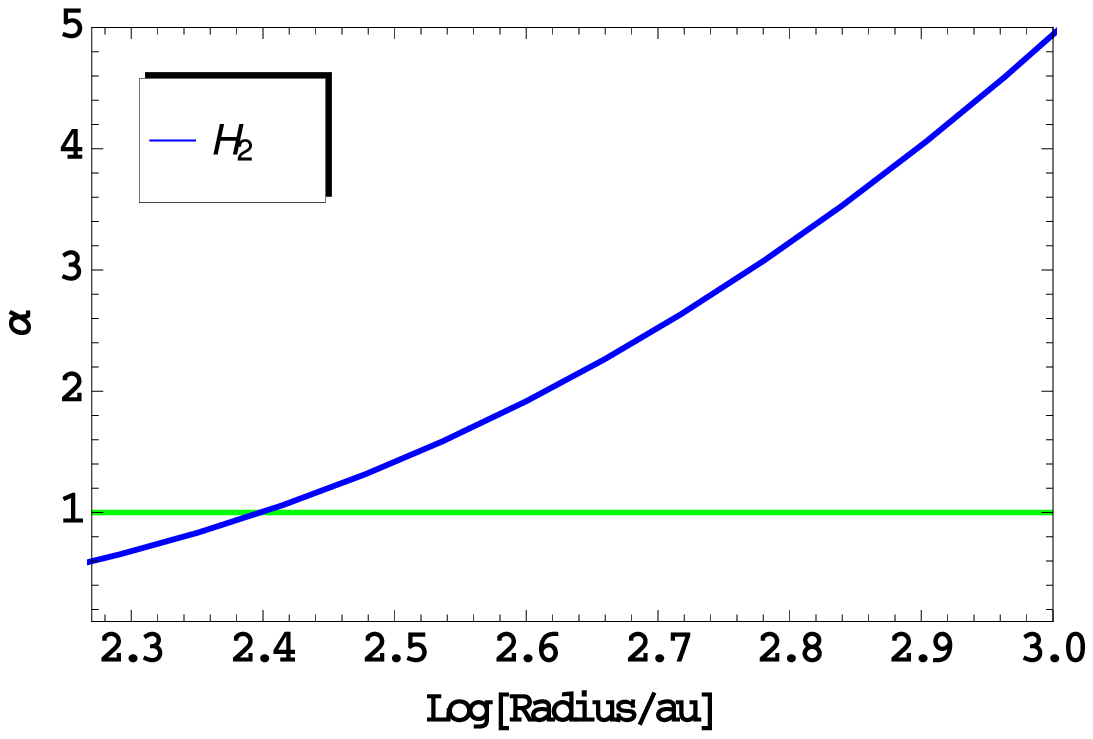}
\end{minipage} \\
\begin{minipage}{6cm}
\vspace{0.2cm}
\includegraphics[scale=0.7]{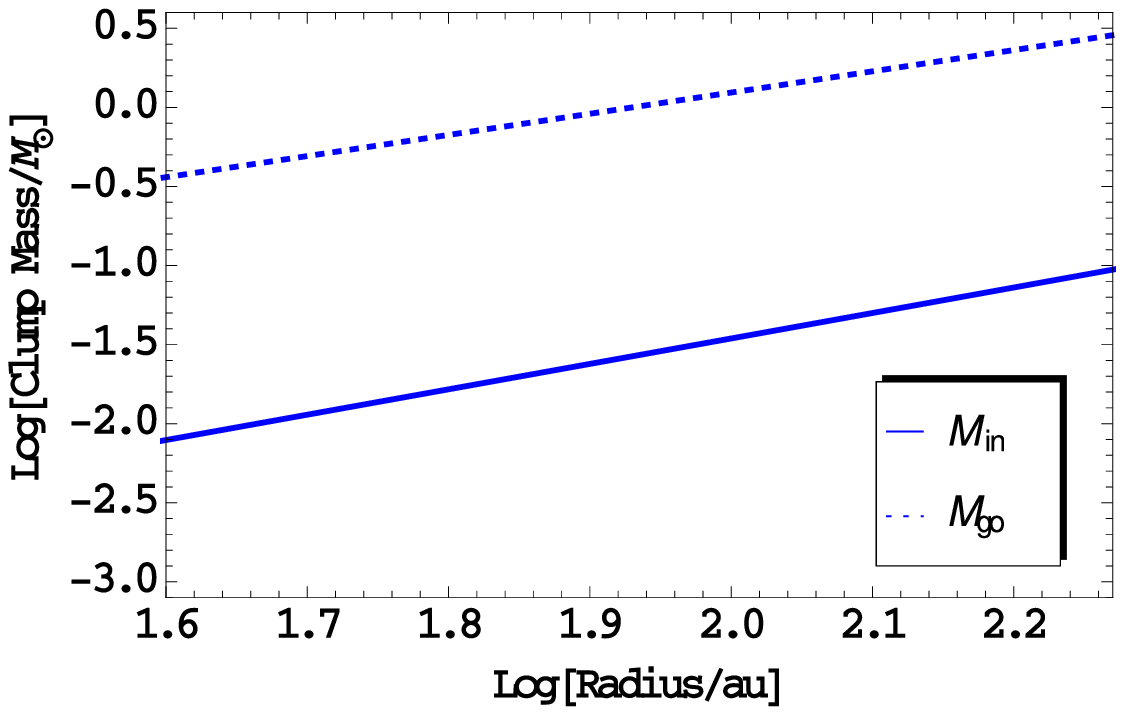}
\end{minipage}&
\begin{minipage}{6cm}
\hspace{2cm}
\includegraphics[scale=0.7]{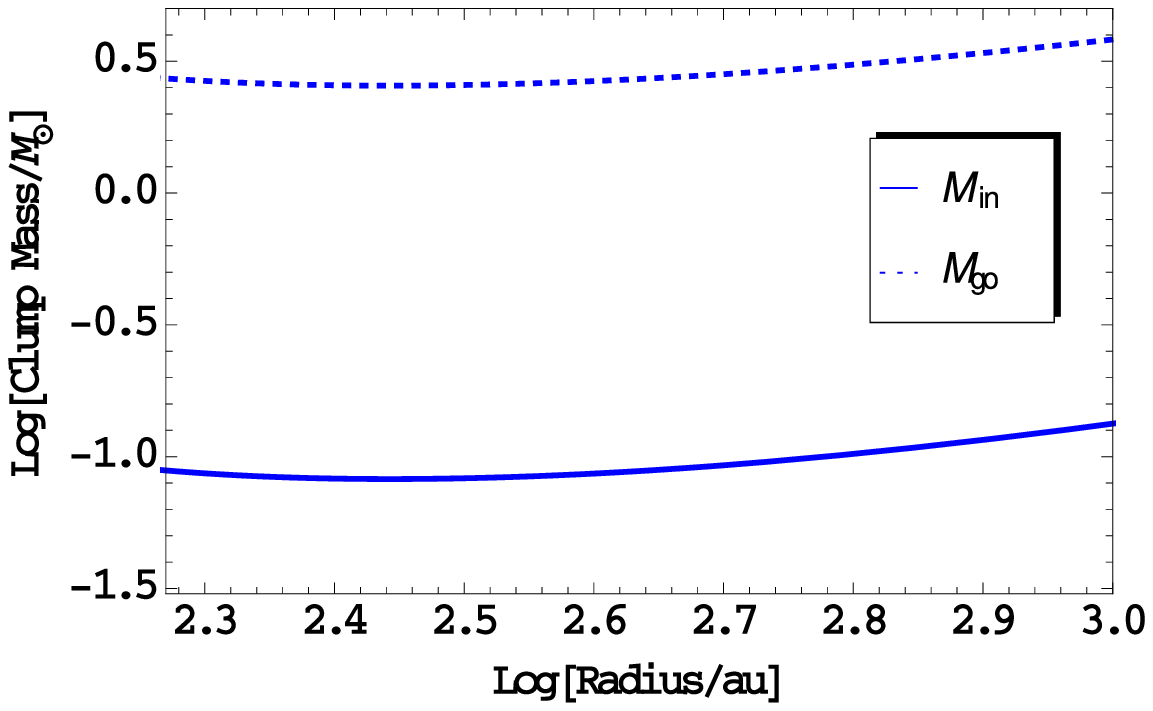}
\end{minipage} 
\end{tabular}
\caption{The disk properties for the central star of $\rm 10^4 ~M_{\odot}$. The temperature, the surface density, the viscous parameter $\alpha$ and the clump masses are shown here. The left panel shows the atomic while the right panel depicts the $\rm H_2$ cooling regime.}
\label{fig4}
\end{figure*}

The analysis of the early and the intermediate stages of the disk evolution shows that the stable radius of the disk continually grows outwards and clumps forming outside this radius are expected to migrate inward. As a result, the central star may reach $\rm 10^4 ~ M_{\odot}$ in about 100,000 years assuming a constant mass supply  of $\rm 0.1~M_{\odot}/yr$. The stellar radius of such a star can be about $40$~AU \citep{Hosokawa13}. Therefore, we adopt the latter as the inner edge of our disk. We have computed the disk properties for a central star of $\rm 10^4~ M_{\odot}$ and show the results in figure \ref{fig4}. It is found that overall the temperature of the disk is increased due to the presence of a very massive central star. The temperature in the disk varies from 1000-4000 K and the atomic cooling regime extends up to a few 100 AU. Overall, the surface density of the disk is about an order of magnitude higher in comparison with the 10 $\rm  M_{\odot}$ case. Our criterion for the stability of the disk shows that the stable radius is about $250$~AU.

\begin{figure*}
\hspace{-6.0cm}
\centering
\begin{tabular}{c c}
\begin{minipage}{6cm}
\includegraphics[scale=0.7]{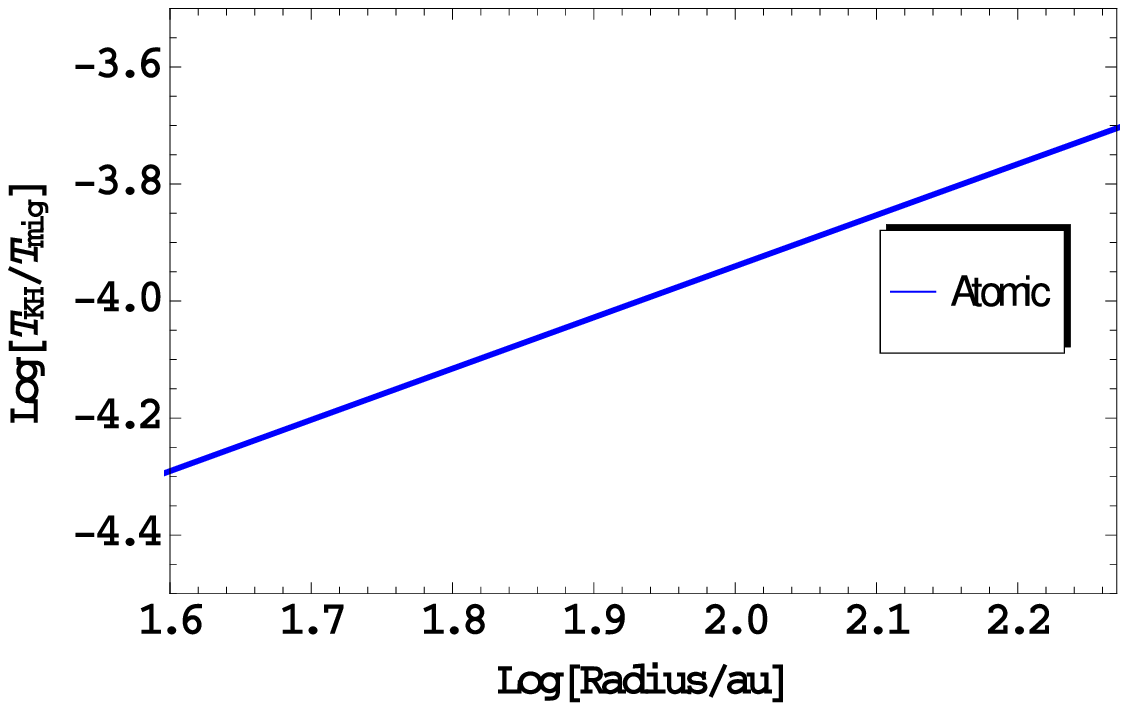}
\end{minipage}&
\begin{minipage}{6cm}
\hspace{2cm}
\includegraphics[scale=0.7]{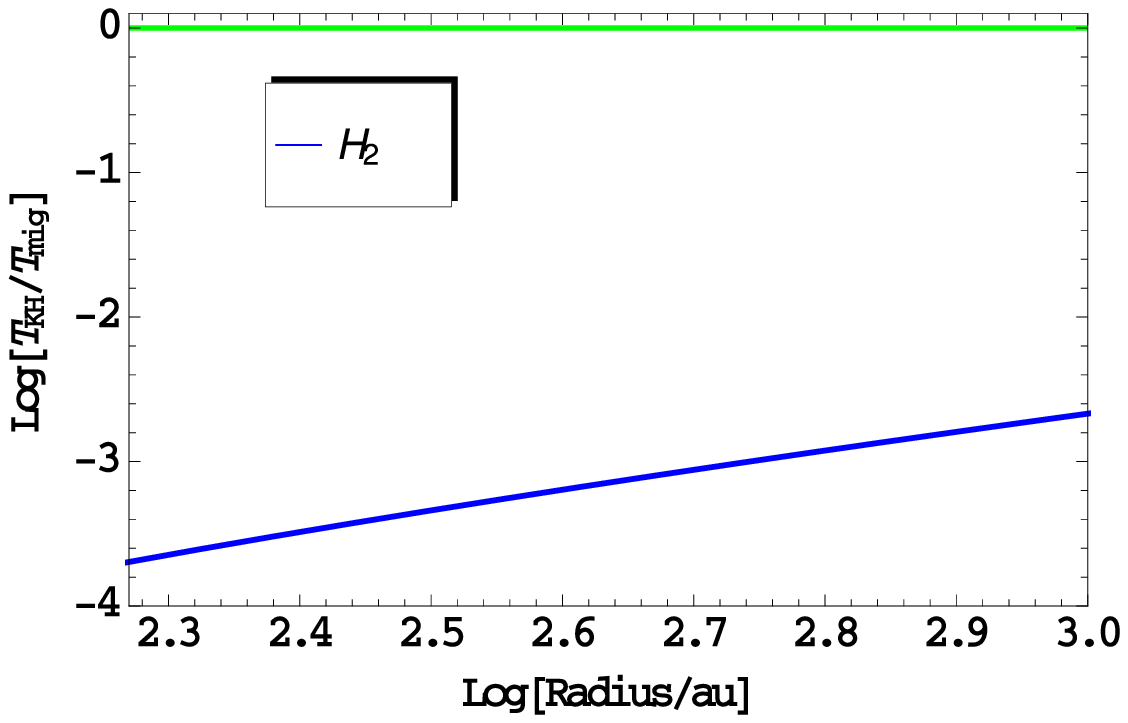}
\end{minipage} \\
\begin{minipage}{6cm}
\vspace{0.2cm}
\includegraphics[scale=0.7]{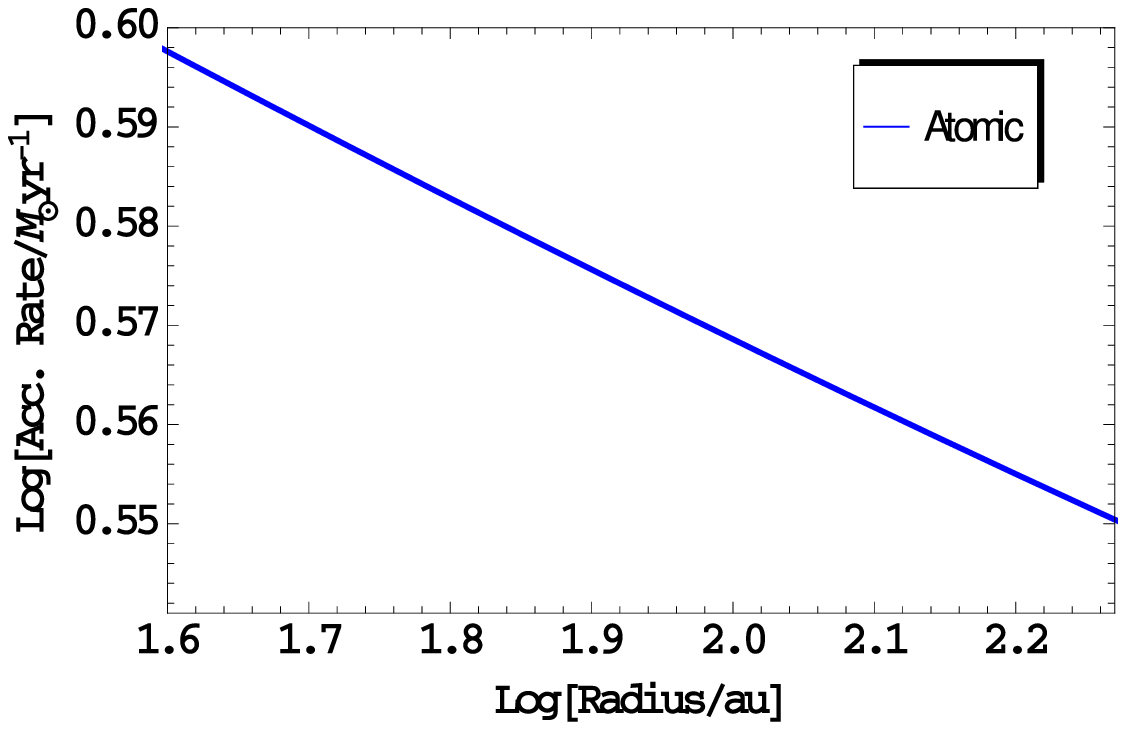}
\end{minipage}&
\begin{minipage}{6cm}
 \hspace{2cm}
\includegraphics[scale=0.7]{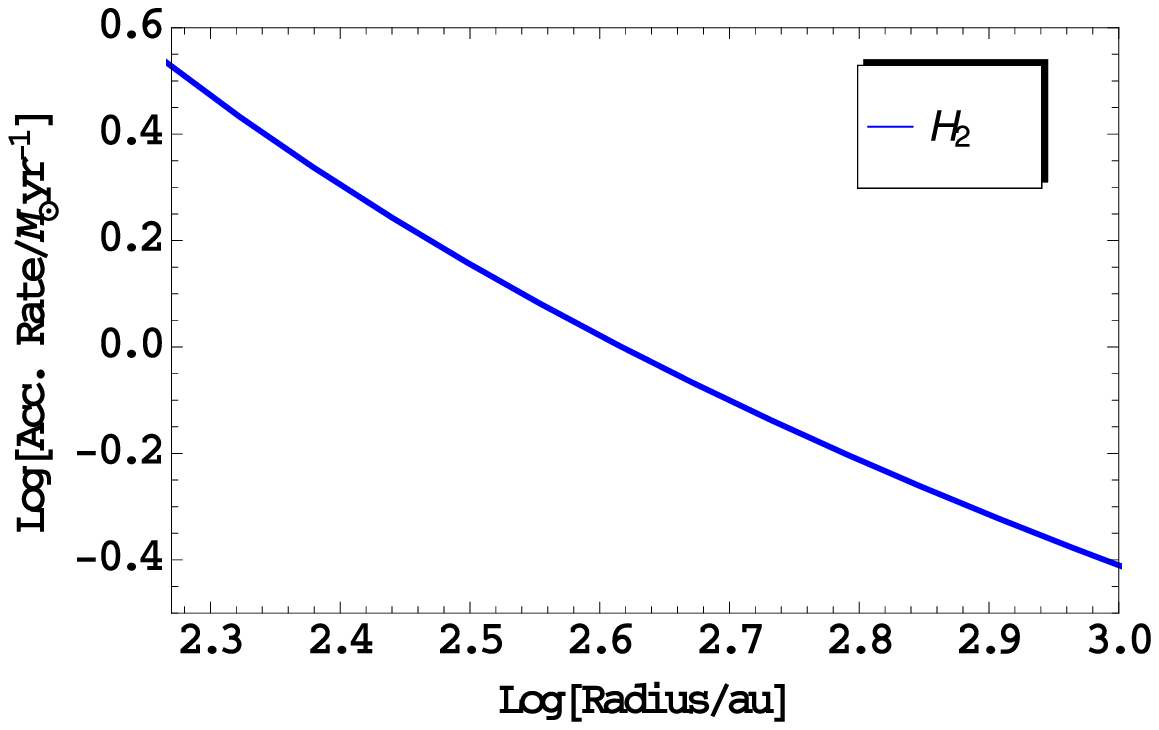}
\end{minipage} \\
\begin{minipage}{6cm}
\vspace{0.2cm}
\includegraphics[scale=0.7]{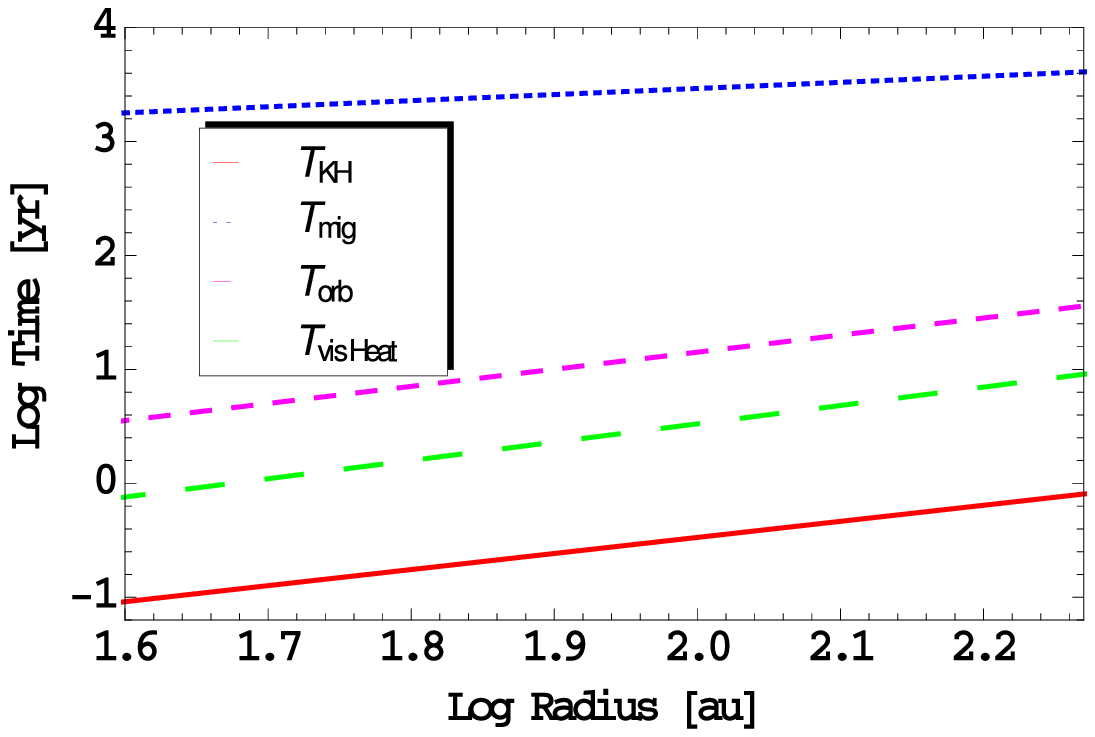}
\end{minipage}&
\begin{minipage}{6cm}
\hspace{2cm}
\includegraphics[scale=0.7]{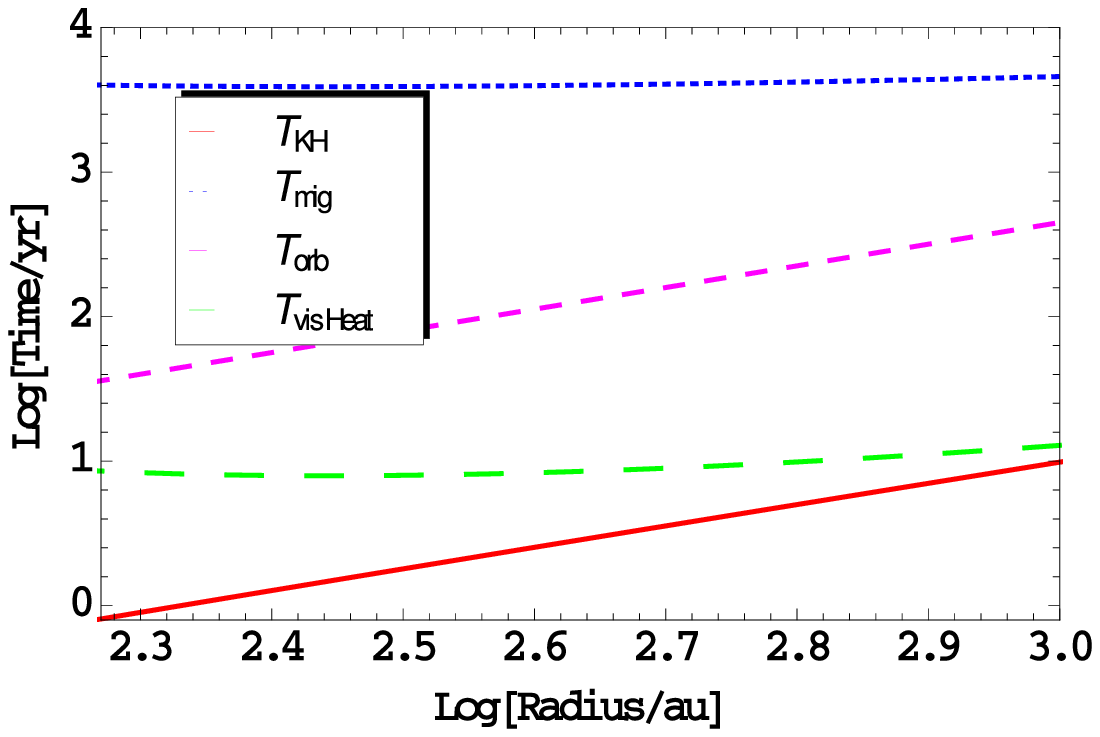}
\end{minipage} 
\end{tabular}
\caption{The disk properties for the central star of $\rm 10^4 ~M_{\odot}$. The migration, the Kelvin-Helmholtz, the orbital and the viscous time scales as well as the accretion rates onto the clumps are shown here. The left panel shows the atomic while the right panel depicts the $\rm H_2$ cooling regime.}
\label{fig5}
\end{figure*}

The masses of the clumps are comparable to the previous cases. The mass accretion rates onto the clumps are almost equivalent to the gas inflow rate onto the disk and even become higher in local regions. In such a case, the accretion onto the clumps will be limited by the gas mass supply to the disk. Our estimates therefore provide an upper limit on the clump accretion rates. The comparison of the migration and the Kelvin Helmholtz time scales is shown in figure \ref{fig5}. The ratio of $\rm T_{KH}$ to $\rm T_{mig}$ drops below one in the unstable regime. It suggests that the clumps may contract towards the main sequence as the Kelvin Helmholtz time becomes shorter due to the higher accretion rates onto the clumps. The comparison of the time scales shows that the viscous heating time again remains shorter compared to the other time scales. We further expect a clumpy mode of accretion in the disk interior due to the migration and disruption of the clumps. For $\alpha_{crit}=1$, the central star is expected to grow further and may reach $\rm 10^{5}~M_{\odot}$ due to continuous gas supply from the stable part of the disk. Our lower limit of the disk photo-evaporation time is about 1000 years, comparable with the migration time scale and much longer than the viscous heating time. It is almost equivalent to the disk accretion timescale.
However, this time represents only a lower limit, and in fact the photo-evaporation may take considerably longer, in particular as rapidly accreting supermassive stars are expected to have cool atmospheres with a temperature of about 5000 K \citep{Hosokawa13,Schleicher13}. Even if that changes at later stages of the evolution, the emission is unlikely to peak exactly at 13.6~eV, therefore decreasing the produced number of ionizing photons. We will therefore assume that photo-ionization will not be relevant during the evolution considered here, but may become relevant at later stages, in particular after the formation of a central black hole.

\begin{figure*}
\hspace{-6.0cm}
\centering
\begin{tabular}{c c}
\begin{minipage}{6cm}
\includegraphics[scale=0.7]{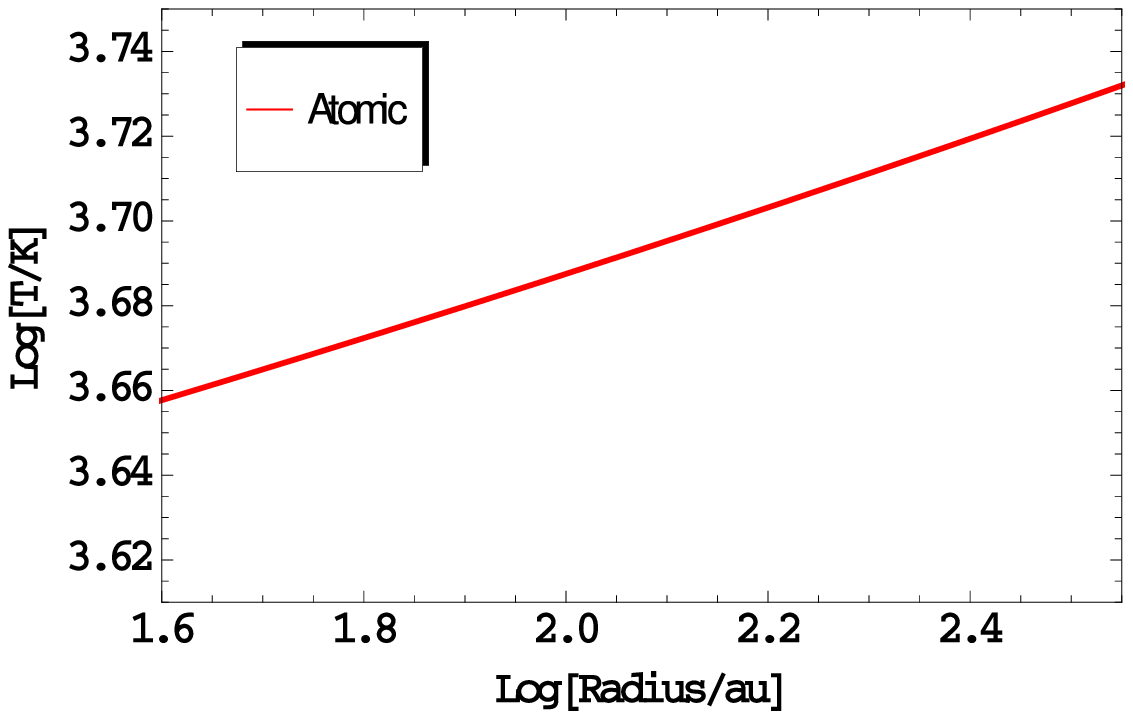}
\end{minipage}&
\begin{minipage}{6cm}
\hspace{2cm}
\includegraphics[scale=0.7]{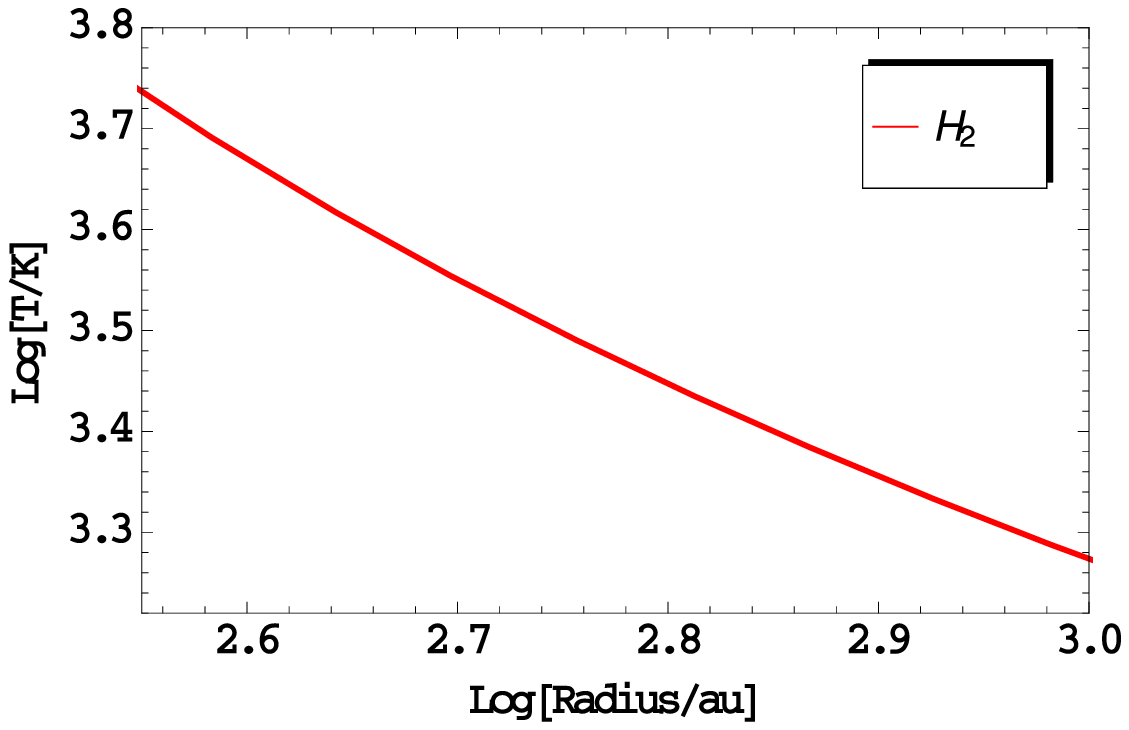}
\end{minipage} \\
\begin{minipage}{6cm}
\vspace{-0.1cm}
\includegraphics[scale=0.7]{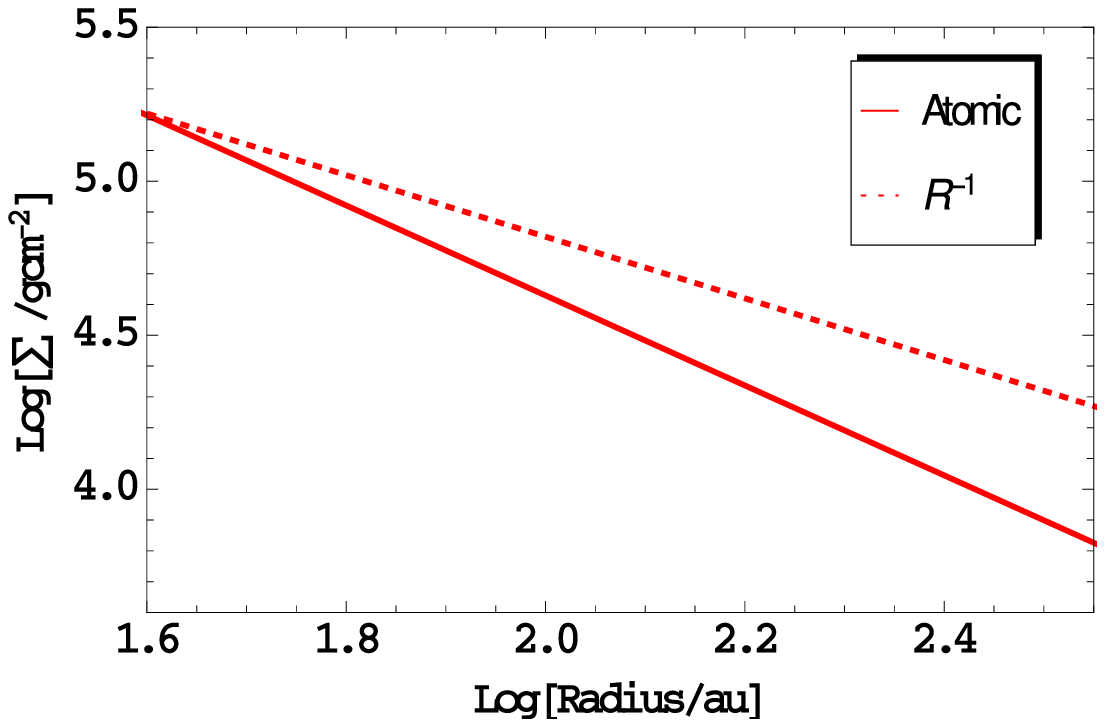}
\end{minipage}&
\begin{minipage}{6cm}
 \hspace{2cm}
\includegraphics[scale=0.7]{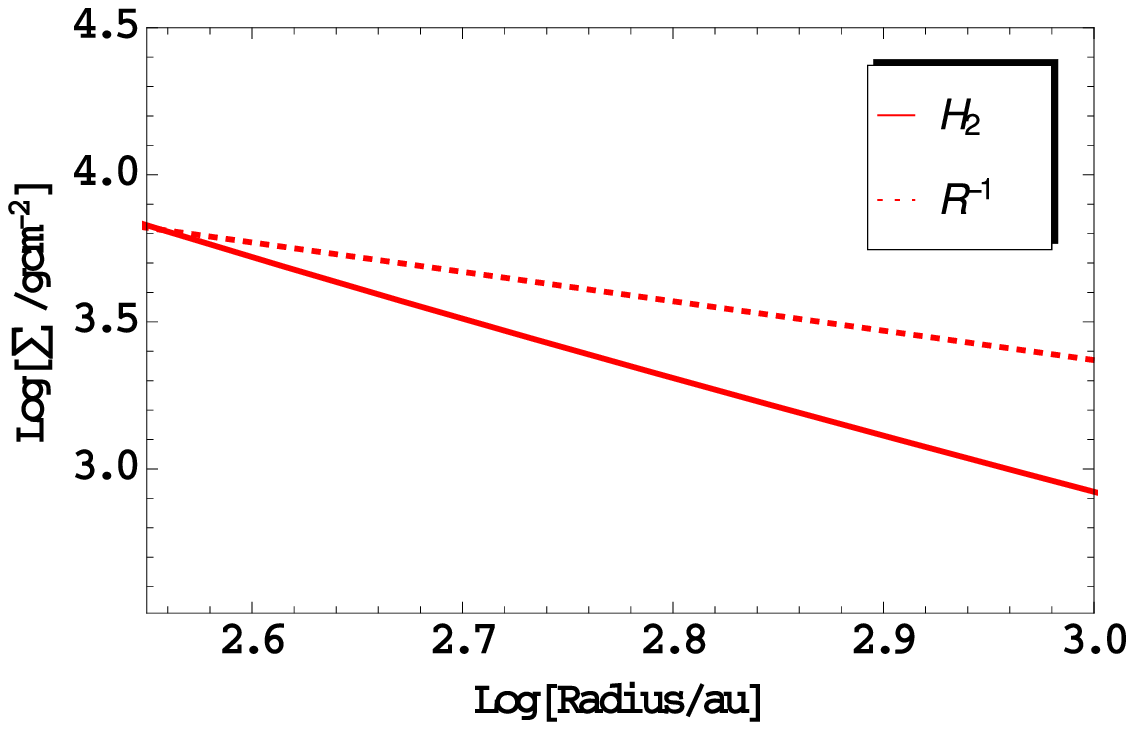}
\end{minipage} \\
\begin{minipage}{6cm}
\includegraphics[scale=0.7]{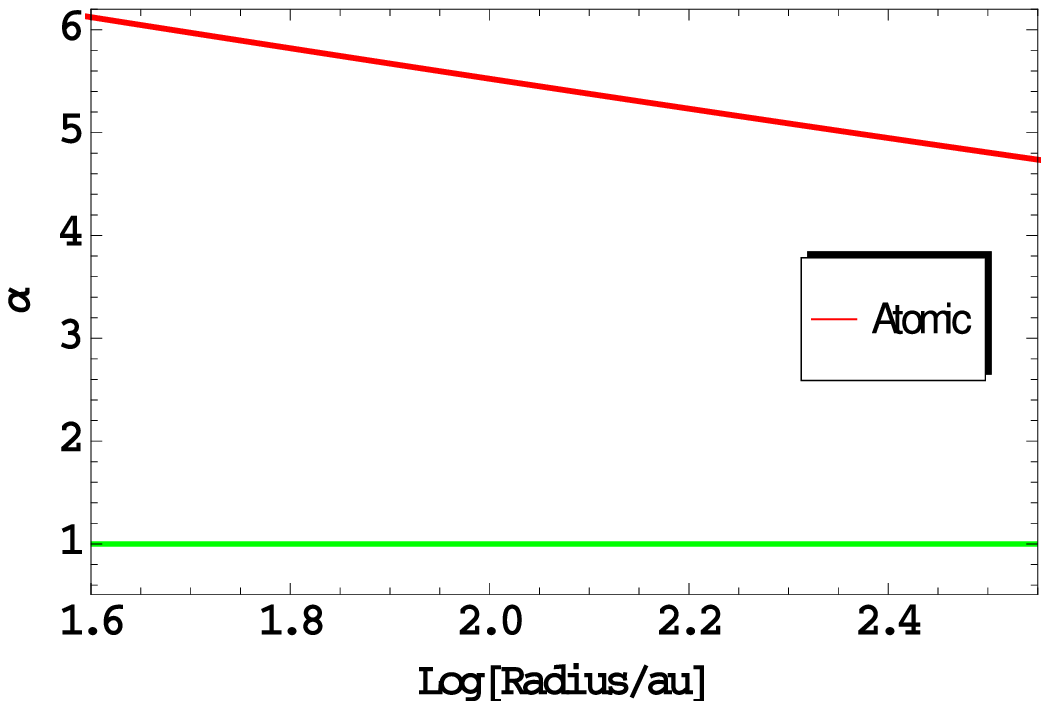}
\end{minipage}&
\begin{minipage}{6cm}
\hspace{2cm}
\includegraphics[scale=0.7]{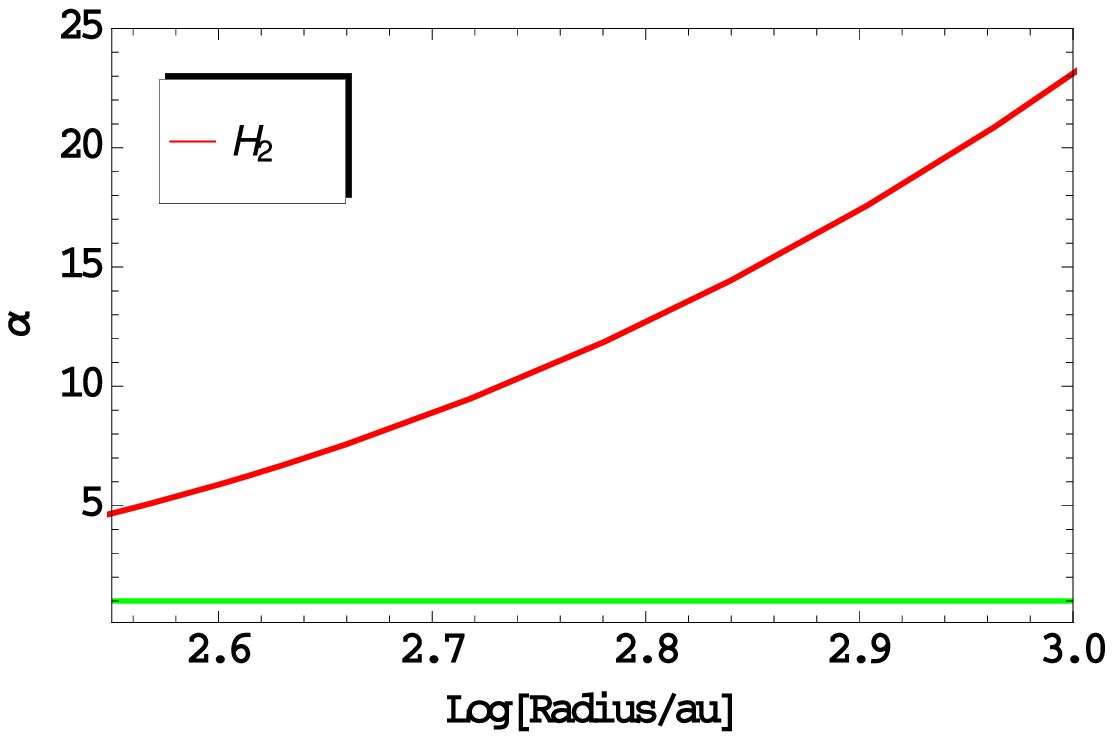}
\end{minipage} \\
\begin{minipage}{6cm}
\vspace{0.2cm}
\includegraphics[scale=0.7]{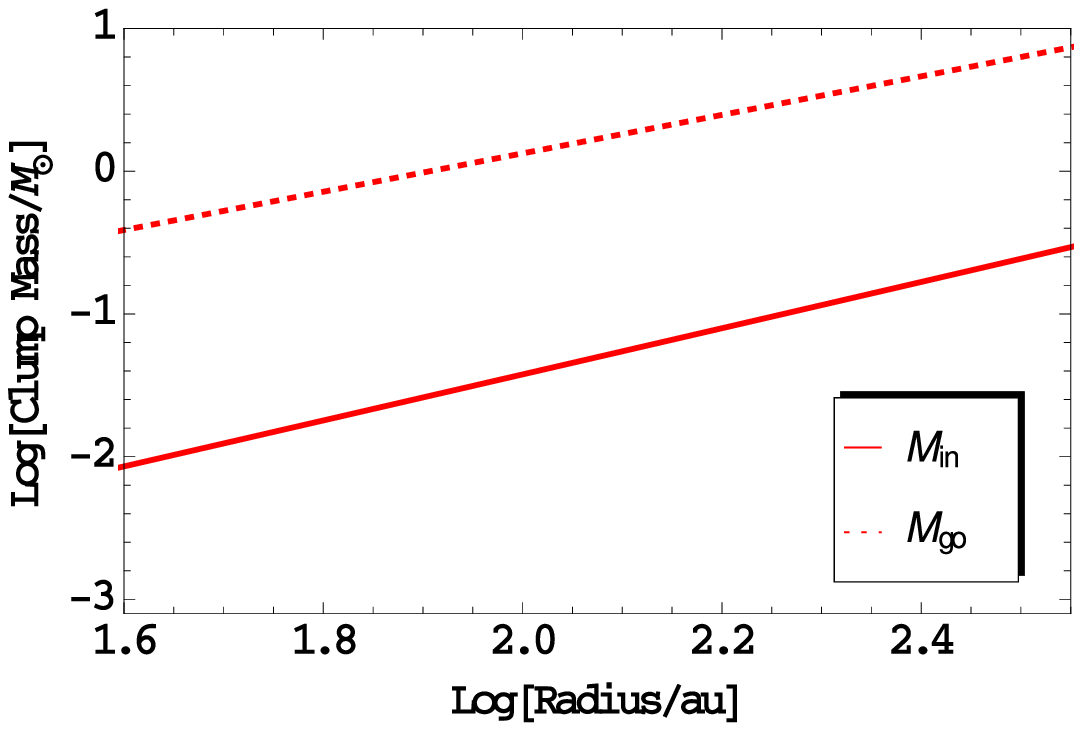}
\end{minipage}&
\begin{minipage}{6cm}
\hspace{2cm}
\includegraphics[scale=0.7]{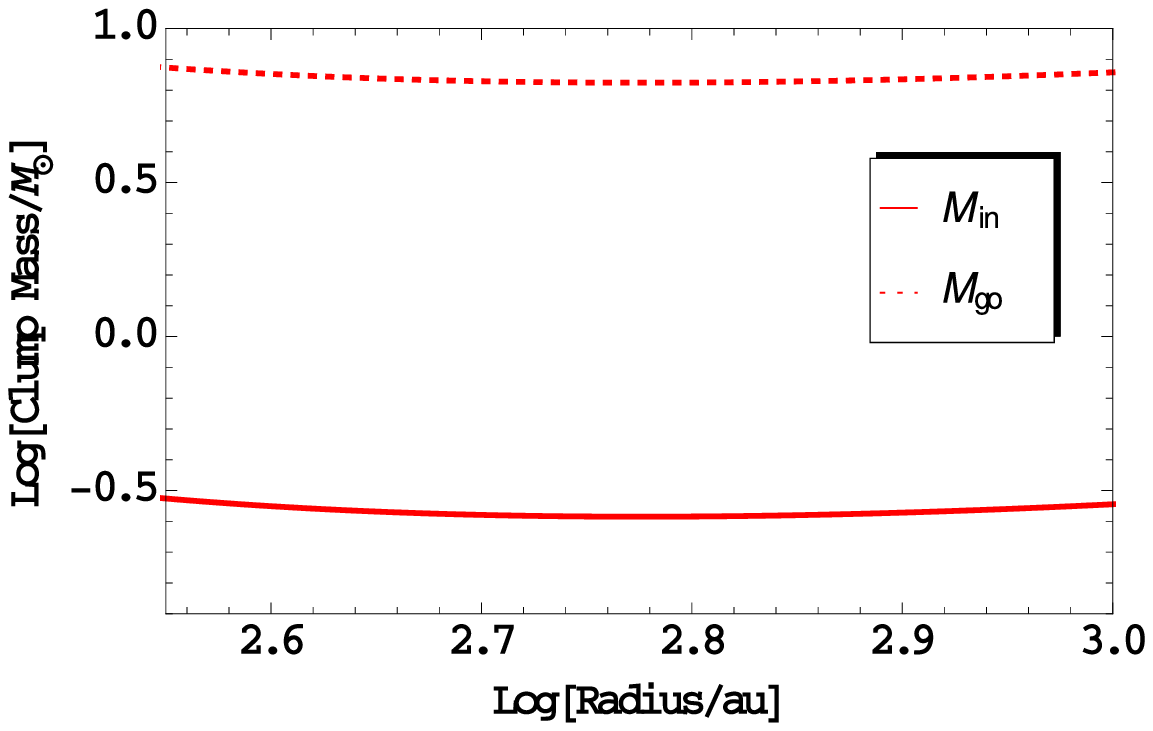}
\end{minipage} 
\end{tabular}
\caption{The disk properties for the central star of $\rm 10^4 ~M_{\odot}$ and the accretion rate of $\rm 1 ~M_{\odot}/yr$. The temperature, the surface density, the viscous parameter $\alpha$ and the clump masses are shown here. The left panel shows the atomic while the right panel depicts the $\rm H_2$ cooling regime.}
\label{fig6}
\end{figure*}

\begin{figure*}
\hspace{-6.0cm}
\centering
\begin{tabular}{c c}
\begin{minipage}{6cm}
\includegraphics[scale=0.7]{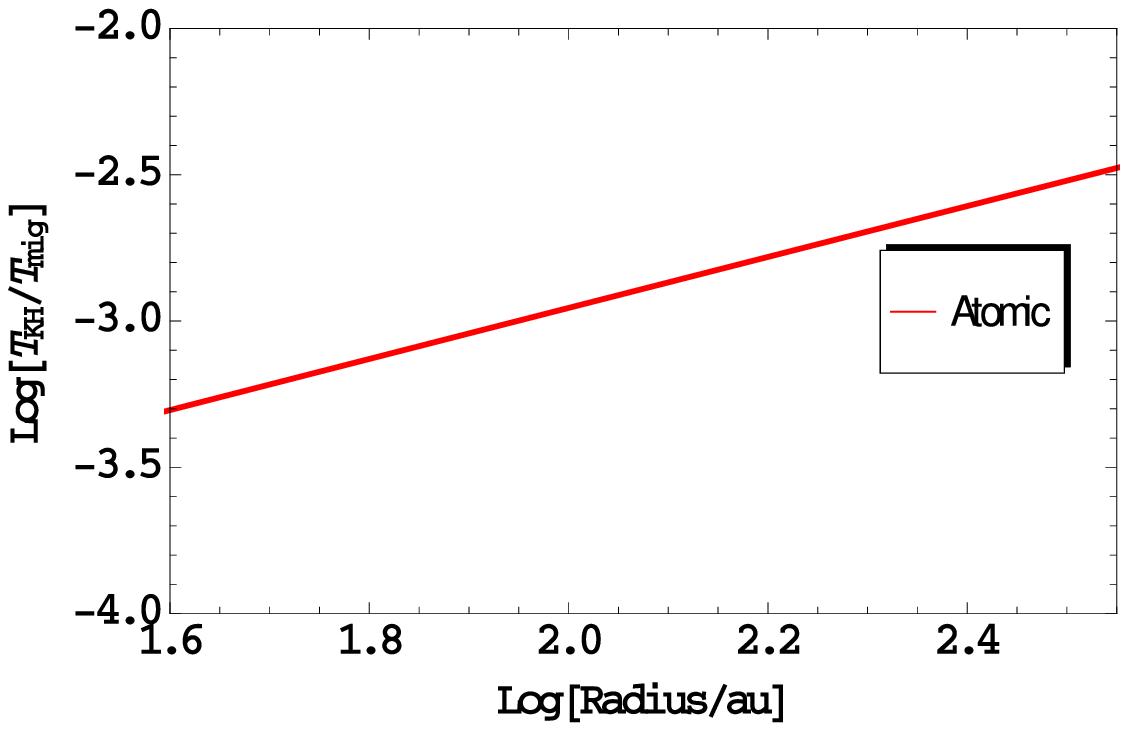}
\end{minipage}&
\begin{minipage}{6cm}
\hspace{2cm}
\includegraphics[scale=0.7]{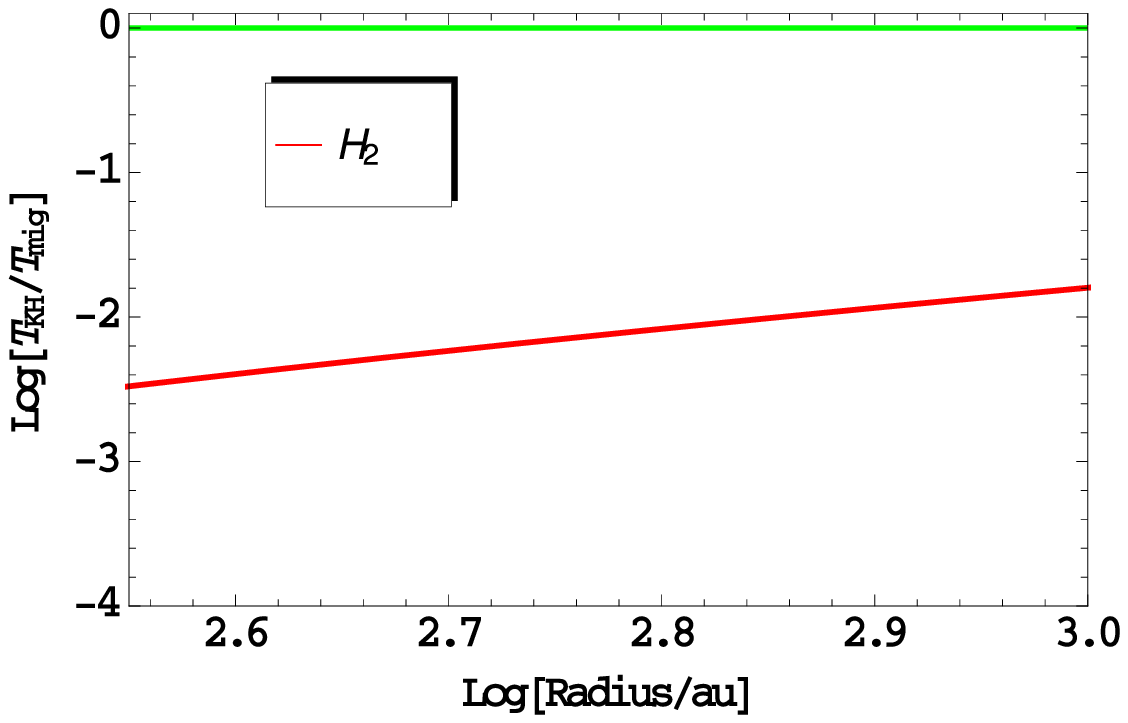}
\end{minipage} \\
\begin{minipage}{6cm}
\vspace{0.2cm}
\includegraphics[scale=0.7]{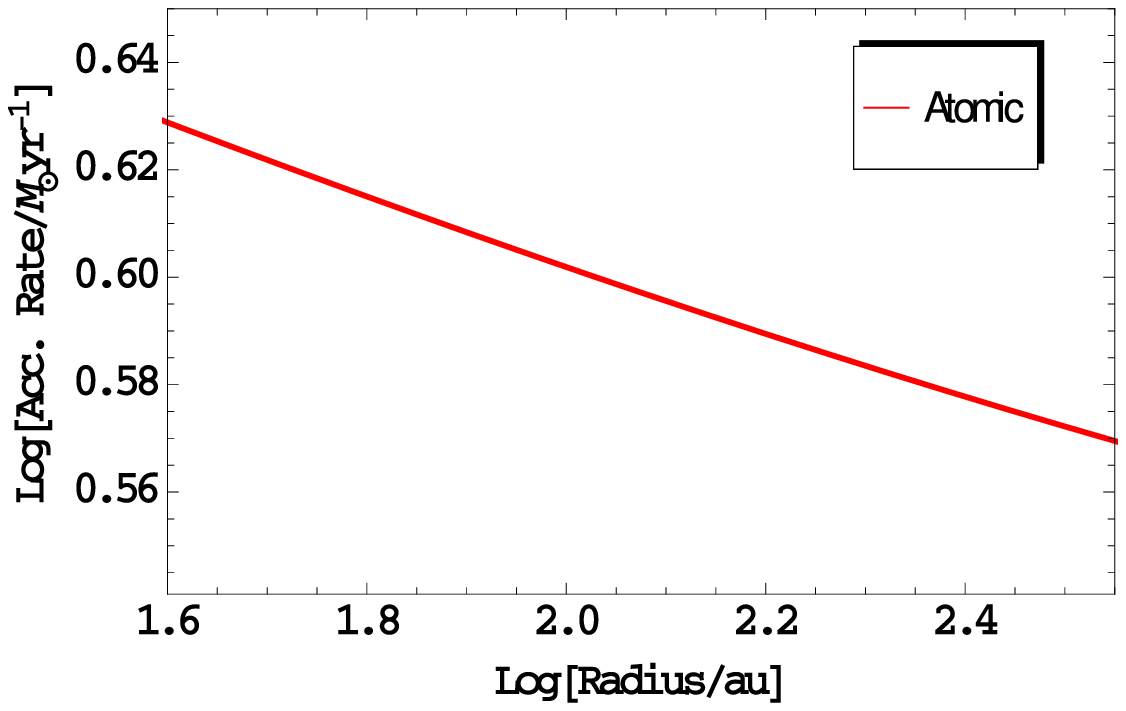}
\end{minipage}&
\begin{minipage}{6cm}
 \hspace{2cm}
\includegraphics[scale=0.7]{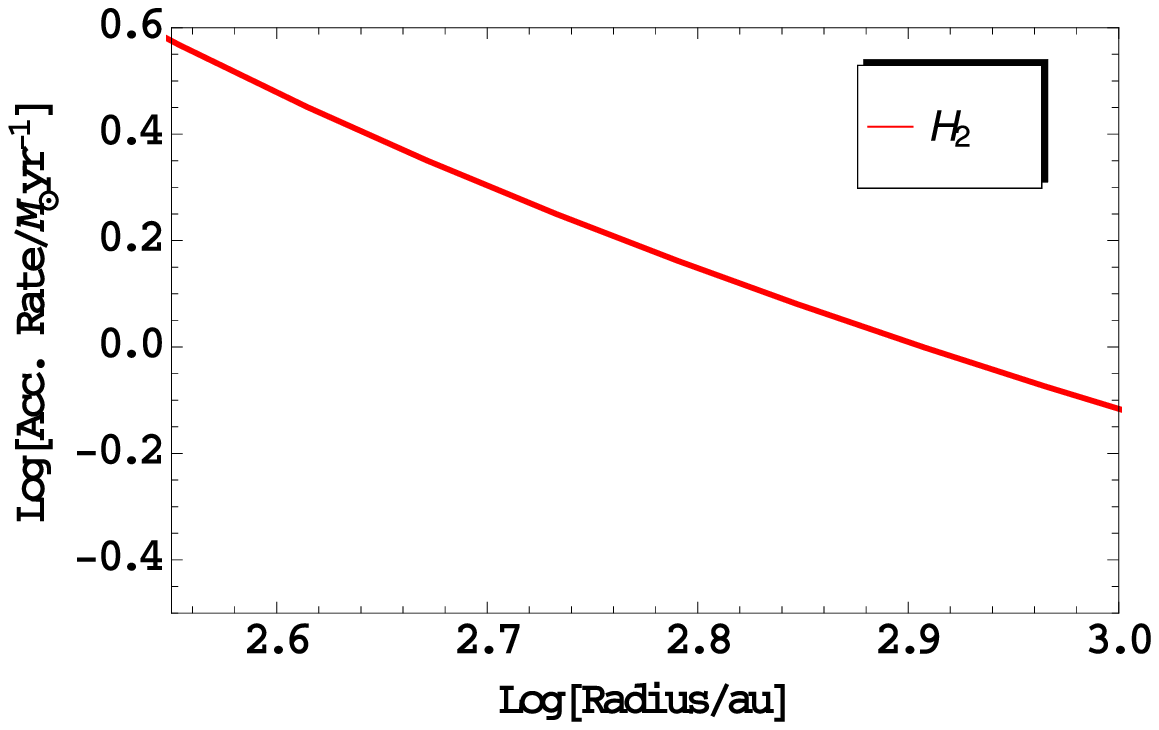}
\end{minipage}\\
\begin{minipage}{6cm}
\vspace{0.2cm}
\includegraphics[scale=0.7]{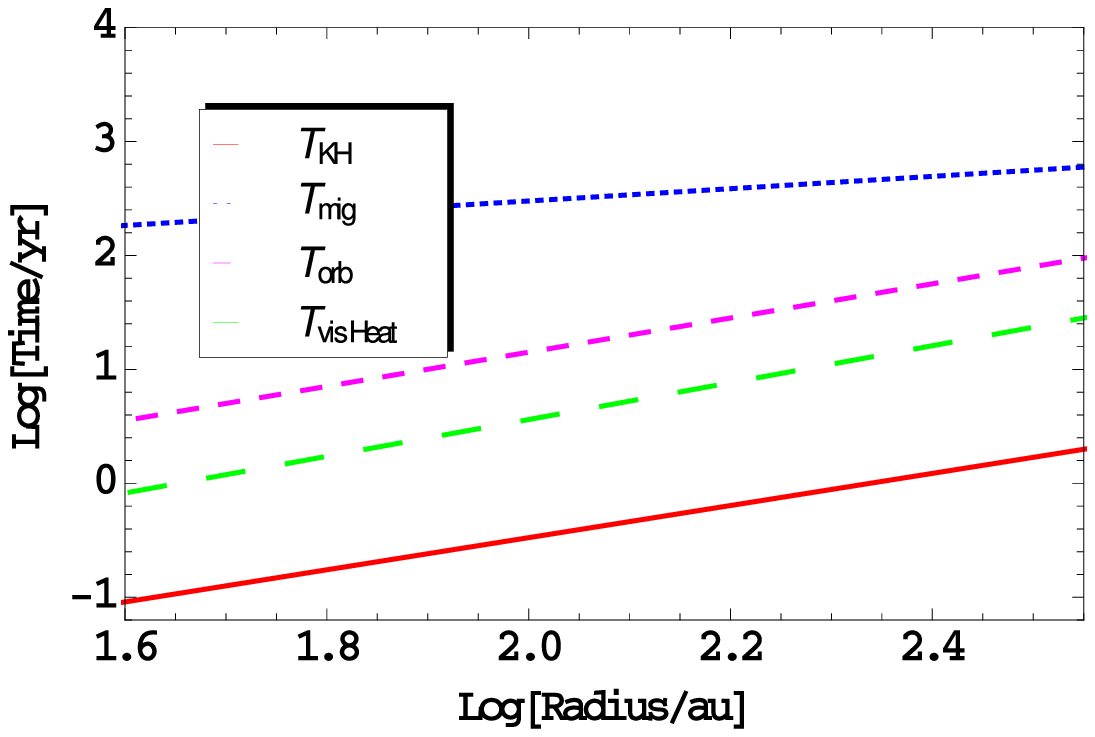}
\end{minipage}&
\begin{minipage}{6cm}
\hspace{2cm}
\includegraphics[scale=0.7]{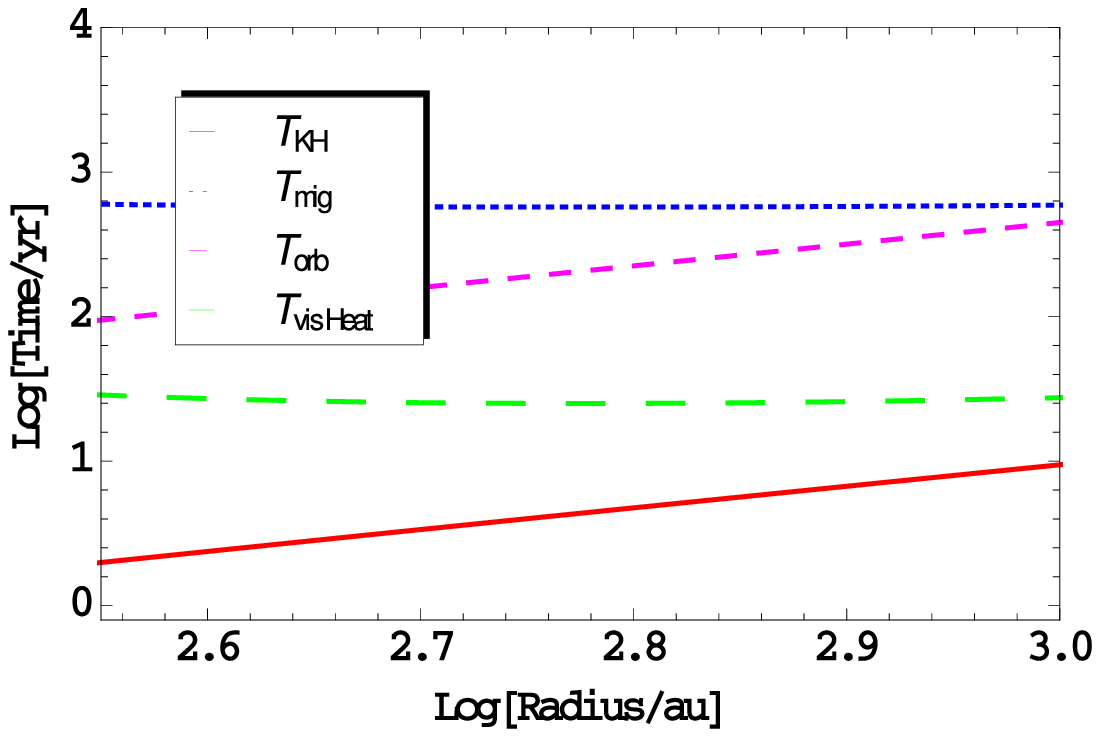}
\end{minipage} 
\end{tabular}
\caption{The disk properties for the central star of $\rm 10^4 ~M_{\odot}$ and the accretion rate of $\rm 1 ~M_{\odot}/yr$.The migration, the Kelvin-Helmholtz, the orbital and the viscous time scales as well as the accretion rates onto the clumps are shown here. The left panel shows the atomic while the right panel depicts the $\rm H_2$ cooling regime.}
\label{fig7}
\end{figure*}

We further explored the case with a mass accretion rate of $\rm 1~M_{\odot}/yr$, as the presence of a very massive central object is expected to increase the accretion rate due to its gravitational attraction, and similar effects have also been found in numerical simulations \citep[e.g.][]{Latif2013d}.  The properties of the disk in this case are shown in figure \ref{fig6}. A notable difference compared to the lower accretion rate is that the disk is even hotter and denser. It is completely unstable, i.e. the value of $\alpha$ is always larger than $1$. The masses of the clumps are further increased by a factor of a few and clumps of about $\rm 10~M_{\odot}$ are expected to form. The mass accretion onto the clumps is comparable to the gas inflow rate to the disk and even increases locally for the reasons outlined above. The comparison of the time scales is shown in figure \ref{fig7}. The ratio of the migration to the Kelvin Helmholtz time scales remains smaller than one similar to the lower accretion rate case (i.e. $\rm 0.1~M_{\odot}/yr$).  In such a case, the disk is highly unstable and vigorous fragmentation will occur. The detailed evolution is difficult to predict in this framework, as the clumps are still exposed to significant accretion, potentially leading to an inward migration. Such a scenario however requires further exploration via numerical simulations.

\subsection{Caveats in our model}\label{uncertainties}
We have assumed the critical value of the viscous parameter to be $\alpha_{\rm crit}$ = 1. This choice is motivated from the findings of numerical simulations of \cite{Zhu2012} employing a detailed treatment of the cooling. However, some studies suggest a critical value of $\alpha_{crit}= 0.06$ \citep{Rice05}, even though their simulations did not consider the detailed modelling of thermal processes. In such a case, the disk in our model will be unstable throughout its evolution and will fragment into multiple clumps. However, the clumps will migrate inward as the migration time remains shorter than the Kelvin-Helmholtz time. During the later stages of the disk evolution in the presence of a very massive ($\rm \sim 10^4~M_{\odot}$) central star, the migration time becomes longer than the Kelvin-Helmholtz time in the interior of the disk and clumps may survive. To improve our current understanding, detailed studies should be pursued to determine the value of $\alpha_{crit}$ under primordial conditions both in the atomic and the molecular cooling regime. 

Our estimate for the Roche limit shows that clumps within the central 10 AU will be tidally disrupted but clumps forming beyond this can exist. If the clumps are able to contract and reach the main sequence, they may emit UV radiation and can inhibit the further inflow of the gas to the center. Our model does not take into account in-situ star formation and their impact on the structure of the disk expected at the later stages of the evolution. Estimates by \cite{Inayoshi2014} have shown that the UV feedback from such stars will not be important at least in the atomic cooling regime, but a further exploration of such processes will be necessary in the future. From the current estimates, it appears very likely that the disk fragments at least on larger scales, leading to the formation of a starburst ring surrounding the central object. The accretion onto the central source may nevertheless continue if enough material falls onto the central parts of the disk, for instance due to the enhanced gravitational potential. At least in some cases, it is conceivable that very massive central objects may form.

We also note that chemical heating and cooling may play an additional role in realistic disk models, in particular at the transition from the molecular to the atomic regime. While this may delay the transition, we note that the viscous heating rate steeply increases with decreasing radius, so that eventually the molecular hydrogen will be completely dissociated. While this may shift the radius where the transition occurs, we expect its impact to be minor due to the steep radial dependence of the viscous heating rate.

Our model is only valid up to the central 1000 AU  of the disk as the assumption of fully molecular disk breaks down on larger scales. If the gas is not fully molecular, the cooling will be somewhat less efficient, and it will be more straightforward for viscous heating to stabilize the gas even on larger scales. In general, the densities in the disk beyond 1000 AU will be lower and we expect  mass accretion rates onto the clumps of about $\rm 10^{-2}-10^{-3}~M_{\odot}/yr$. This will result in a Kelvin Helmholtz time of about $\rm 10^{5}~yr$, and $10^4$~M$_\odot$ can be accreted before feedback on these scales becomes relevant. 

Finally, the clump migration presents an additional uncertainty in the disks considered here, and it can be potentially enhanced if the disks are very clumpy. The latter may increase the effect of migration in particular at the late stages of the evolution. At the same time, three-body effects may play a role and lead to the ejection of some of the clumps, thus  reducing the accretion onto the central source.

\section{Discussion and conclusions}

We present here a simple model for a disk forming in a massive primordial halo of $\rm 10^7-10^8~M_{\odot}$. The halo is irradiated by a moderate  Lyman-Werner flux and high accretion rates of about $\rm 0.1~M_{\odot}/yr$ are expected to occur. We explore  various stages of the disk evolution with a central object of 10, 100 and $\rm 10^4~M_{\odot}$. It is presumed that the disk is fully molecular, in a steady state condition and metal-free. The properties of the disk are computed assuming that the disk is in thermal equilibrium, i.e. the cooling is balanced by the viscous heating due to the turbulence and spiral shocks. We presume that the disk becomes unstable for the critical value of the viscous parameter $\alpha = 1$. This study enables us to quantify the effect of viscous heating and helps us to understand  the formation of a massive central object  in such a disk. These considerations hold both in the presence of a massive central star as well as in the presence of a massive black hole.

Our results show that the disk is initially cooled by molecular hydrogen, which later gets dissociated as viscous heating dominates and a transition from the molecular to the atomic cooling takes place. The temperature in the disk interior remains about 3000-4000 K. The role of viscous heating becomes more important during the later stages of the disk evolution as the central star becomes more massive. In this case, the range of the atomic cooling extends outwards. For the case of a central star of $\rm 10^4~M_{\odot}$, the size of the atomic cooling regime is about $200$~AU. Overall, as the central star becomes more massive, the disk becomes denser and hotter. Our criterion for the stability of the disk (i.e. $\alpha_{crit} = 1$) shows that the disk remains stable in the interior and becomes unstable in the outskirts. For a central star of $\rm 10~M_{\odot}$, the size of the stable radius is $30$~AU. It extends to $300$~AU for a central star of $\rm 10^4~M_{\odot}$. The disk is expected to fragment beyond the stable radius. Clumps with masses of a few M$_\odot$ are expected to form in the unstable regime. The accretion rates onto the clumps become higher as the central star becomes more massive.
  
To assess the fate of these clumps, we have computed the migration and the Kelvin-Helmholtz time scales. The comparison of the time scales shows that the migration time remains much shorter than the Kelvin-Helmholtz time  during the early stages of the disk evolution and the clumps forming in the unstable regime are migrated inward. However, as the central star becomes more massive ($\rm \sim10^4$~M$_\odot$), an inversion of the time scales takes place outside the stable radius. The Kelvin-Helmholtz time  then becomes shorter than the migration time scale. In this case, the clumps will contract and evolve towards the main sequence. The estimates of the Roche limit show that in this case the clumps forming within the central $10$~AU will be tidally disrupted. Therefore, the central clump may further reach $\rm 10^5~M_{\odot}$ for $\alpha_{crit}$=1 if there is still a sufficient gas supply to the stable disk, for instance due to a more direct infall, which can be enhanced due to the mass of the central object. For $0.06 < \alpha_{crit} < 1$, the disk becomes unstable on all scales, therefore enhancing fragmentation. In this case, maintaining the accretion onto the central clump may be more difficult. An accurate determination of $\alpha_{crit}$ is therefore important particularly in this regime.


We also investigated a case with an accretion rate of $\rm 1~M_{\odot}/yr$ for a central star of $\rm 10^4~M_{\odot}$. We find that the disk becomes highly unstable in this case and the migration time scale also remains longer than the Kelvin-Helmholtz timescale. In such a case, strong fragmentation may occur. A final conclusion on this scenario is however not straightforward, due to the continuing and ongoing accretion. It will therefore be important to assess the evolution of the late phases of such disks also with numerical simulations. A potential concern for models of accretion disks is the presence of outflows, which may reduce the accretion onto the central object. Such outflows may be driven either via radiative feedback, or as a result of magnetic processes in the accretion disk. Supermassive stars forming  via rapid mass accretion of the order of $\rm \geq 0.01~M_{\odot}/yr$ do not produce strong UV feedback due to their low surface temperatures. Thus, no strong outflow powered by radiation feedback is expected from these stars. However, it is conceivable that the inner part of the disk is efficiently magnetized via dynamos, potentially leading to the formation of jets and outflows (\cite{Tan2004}; Latif \& Schleicher, in prep.). 


We have also presumed that the disk is metal free, but this may not always be the case. The halos irradiated by UV flux may also get polluted due to supernova winds. If there are trace amount of dust and metals, they can cool the gas to lower temperatures \citep{2005ApJ...626..627O,2009A&A...496..365C,2012A&A...540A.101L,Bovino14}. Although we still expect the viscous heating to be dominant within the disk interior, metals and dust are likely to enhance the instability on larger scales. In such a case, a stellar cluster or a starburst ring may form on larger scales surrounding the central object, and its evolution in the presence of ongoing inflows needs to be explored in future studies. 

Overall, our results suggest that very massive objects may form even in the presence of a moderate UV background, and that viscous heating may play a central role in stabilising the disks surrounding these stars.  Therefore, massive black holes of about $\rm 10^5~M_{\odot}$ may form  in rapidly rotating  metal free accretion disk exposed to a large inflow rate of $\rm 0.1 ~M_{\odot}/yr$.  Our study further confirms the hypothesis that a complete isothermal collapse is not necessary to form a direct collapse black hole, and alleviates the need for an extremely strong radiation background in this scenario. We conclude that a moderate background may be sufficient to maintain high accretion rates on large scales, while viscous heating in the center may stabilize the gas in the center during the formation of massive objects. As a result, the proposed mechanism will considerably enhance the abundance of massive black holes at $z \sim$ 10. This has important implications for the expected number density of direct collapse black holes observed at $z \geq$ 6.

\section*{Acknowledgments}
We thank Marta Volonteri, Andrea Ferrara and Stefano Bovino for interesting discussions and helpful suggestions. The research leading to these results has received funding from the European Research Council under the European Community's Seventh Framework Programme (FP7/2007-2013 Grant Agreement no. 614199, project ``BLACK''). DRGS is indebted to the Scuola Normale Superiore (SNS) in Pisa for an invitation as a 'Distinguished Scientist', which has stimulated the research described in this manuscript.

\bibliography{disk.bib}

\end{document}